\documentclass[11pt,reqno]{article}
\usepackage[lmargin=2cm, rmargin=2cm]{geometry}
\usepackage[dvips]{graphicx}
\usepackage{amssymb}
\usepackage{appendix}
\usepackage[utf8]{inputenc}
\usepackage[english]{babel}
\usepackage{amsmath, amssymb, graphics}
\usepackage{amsfonts, color}
\usepackage{mathrsfs}
\usepackage{cite}

\allowdisplaybreaks \numberwithin{equation}{section}

\newcommand{\mathsym}[1]{{}}

\newcommand{\mc}{\mathcal}

\newcommand{\p}{\partial}

\newcommand{\rar}{\rightarrow}

\newcommand{\nn}{\nonumber}
\newcommand{\im}{\mathrm{Im}}

\def\bfone{\relax{\rm 1\kern-.35em 1}}

\makeatletter \@addtoreset{equation}{section} \makeatother

\usepackage{hyperref}

 \allowdisplaybreaks[3]

\begin{document}

\begin{titlepage}
 \thispagestyle{empty}
 \begin{flushright}
     \hfill{ITP-UU-14/29 }\\
       \hfill{CPHT-RR098.1214 }\\
 \end{flushright}

 \vspace{55pt}

 \begin{center}
     { \LARGE{\bf      {Rholography, Black Holes and Scherk-Schwarz}}}

     \vspace{50pt}

\Large{Nava Gaddam$^1$, Alessandra Gnecchi$^2$, Stefan Vandoren$^1$, Oscar Varela$^{3,4}$} \\[8mm]
{\small\slshape
${}^1$ Institute for Theoretical Physics \emph{and} Center for Extreme Matter and Emergent Phenomena, \\
Utrecht University, 3508 TD Utrecht, The Netherlands \\[3mm]

${}^2$Institute for Theoretical Physics, KU Leuven, 3001 Leuven, Belgium\\[3mm]

${}^3$ Center for the Fundamental Laws of Nature, \\ Harvard University, Cambridge, MA 02138, USA \\[3mm]

${}^4$ Centre de Physique Th\'eorique, Ecole Polytechnique, CNRS UMR 7644, \\ 91128 Palaiseau Cedex, France

\vspace{5mm}

{\upshape\ttfamily gaddam@uu.nl, alessandra.gnecchi@fys.kuleuven.be, s.j.g.vandoren@uu.nl, ovarela@physics.harvard.edu}\\[3mm]}

\vspace{8mm}

     \vspace{10pt}

    \vspace{10pt}

     {\bf Abstract}
     \end{center}

We present a construction of a class of near-extremal asymptotically flat black hole solutions in four (or five) dimensional gauged supergravity with R-symmetry gaugings obtained from Scherk-Schwarz reductions on a circle. The entropy of these black holes is counted holographically by the well known MSW (or D1/D5) system, with certain twisted boundary conditions labeled by a  twist parameter $\rho$. We find that the corresponding $(0,4)$ (or $(4,4)$) superconformal algebras are exactly those studied by Schwimmer and Seiberg, using a twist on the outer automorphism group. The interplay between R-symmetries, $\rho$-algebras and holography leads us to name our construction ``Rholography".

 \vspace{10pt}
\noindent

\end{titlepage}


\thispagestyle{plain}

\tableofcontents

\baselineskip 6 mm

\section{Introduction and Summary}

Among the successes of string (M-) theory, the interpretation of black hole entropy as resulting from microscopic degrees of freedom of D-branes (M-branes) stands out. Since the inspirational work\textemdash valid for BPS black holes\textemdash of \cite{Strominger:1996sh,Maldacena:1997de}, attempts at understanding non-supersymmetric and non-extremal black holes have also been initiated \cite{Callan:1996dv,Horowitz:1996fn,Breckenridge:1996sn,Dabholkar:1997rk,Sen:2007qy}. 
The black hole solutions described in \cite{Strominger:1996sh,Maldacena:1997de} can be uplifted to black strings in one higher dimension, where the near-horizon limit contains an $AdS_3$ factor in the supersymmetric case, or a BTZ-factor for near-extremal black holes. One can then use the Cardy formula for the dual conformal field theories, which are based on $(0,4)$ \cite{Maldacena:1997de} and $(4,4)$ \cite{Strominger:1996sh} superconformal field theories in two dimensions. In the near-extremal case, one makes use of the correspondence between BTZ geometries and thermal conformal field theories, which does not rely on any supersymmetry \cite{Strominger:1997eq}.\\

Among the plethora of asymptotically flat black holes whose microscopics have been studied, there have been none in gauged supergravity, in which there is light charged matter in the supergravity spectrum. Such a situation would be needed to study absorption and reflection coefficients of charged matter by the black hole, or to compute black hole discharge through Schwinger processes. In this article, we will not go as far as this, but we present a construction of a class of black hole solutions in gauged supergravity with flat Minkowski vacua. In other words, we present a framework in which these processes could be studied. 
Microscopically, we find that they are described by $N=4$ superconformal systems ($(0,4)$ or $(4,4)$) with twisted boundary conditions, characterized by a parameter, called $\rho$ in the original work of Schwimmer and Seiberg \cite{Schwimmer:1986mf}. The twist involves the outer automorphism group of the superconformal algebra, and the corresponding $\rho$-algebra is inequivalent to the usual NS and R-sectors and its spectral flow \cite{Schwimmer:1986mf}.
To the best of our knowledge, this $\rho$-algebra has never found an interesting application, but in this paper we show that it governs the microscopics of asymptotically flat black holes in gauged supergravities.\\

We will first concern ourselves with M-theory on a compact Calabi-Yau threefold
($CY_3$, henceforth). This results in five-dimensional $\mathcal{N}=2$ supergravity coupled to vector- and hypermultiplets. The low-energy limit of the M5-brane put on a compact divisor in the $CY_3$ is a black string solution to the said supergravity theory. A further compactification of this black string along an $S^1$ is a black hole solution of a four-dimensional $\mathcal{N}=2$ supergravity theory. As was shown in \cite{Maldacena:1997de}, the macroscopic entropy of such a BPS black hole is microscopically realized as a set of microstates within the conformal field theory living on the worldsheet of the M5-string.
This theory has been called the MSW-CFT in the literature; it is a (0,4) superconformal field theory (SCFT) in 2 dimensions. The M5-string worldsheet effective action has been studied in detail, in \cite{Minasian:1999qn} for instance.\\

While we will still persist with five-dimensional ungauged supergravity theory resulting from compactifications of M-theory, we wish now to consider a more general Scherk-Schwarz reduction along the additional $S^1$ to arrive at a four-dimensional theory. Such a consideration is not new. In \cite{Andrianopoli:2004im,Looyestijn:2010pb}, it was shown that imposing Scherk-Schwarz twisted boundary conditions  results in gauged supergravities theory in four dimensions with positive definite scalar potentials with Minkowski vacua. Our strategy will be to find a Scherk-Schwarz twist and the corresponding gauging that preserves the black hole solutions from the untwisted case. As we will show, such a twist can be done by using the R-symmetry group. R-symmetry in supergravity is in general not a symmetry of the action, but classically - ignoring quantum corrections - it often is. We can say it is an approximate symmetry, valid in the classical supergravity regime, and use it in the Scherk-Schwarz twist. Since the theory is now gauged, the spectrum is non-trivial and there are light, R-charged particles in it. Owing to the fact that the circle on which we Scherk-Schwarz reduce is exactly the spatial circle of the M5-string, boundary conditions can consistently be imposed in the microscopics to match the macroscopic supergravity setup. In essence, a twisted generalization of the MSW-CFT results. The interplay between R-symmetry, the $\rho$-twist and the use of holography motivates us to name our construction ``R-Holography", ``$\rho$-lography", or simply ``Rholography".\\

In Section \ref{sec:rhoalgebras}, we review the $\rho$-algebras of \cite{Schwimmer:1986mf} and show how the boundary conditions are implemented in the bulk on the gravitini living in $AdS_3$. Moreover, we will study the implications of the $\rho$-twist for the ground state energy of the right moving sector of the MSW-CFT. This will allow us to determine the entropy of the field theory in a thermal state, as a function of $\rho$, using Cardy's celebrated formula. \\

Further on, in Section \ref{sec:bulk}, we will identify the appropriate Scherk-Schwarz twist in the five-dimensional supergravity theory that corresponds to the microscopic $\rho$-twist and carry out the reduction to four dimensions. As we will show, the relevant twist uses the R-symmetry that acts on the supersymmetry generators of the five-dimensional $\mathcal{N}=2$ supergravity theory. After identifying a flat Minkowski vacuum we compute the supergravity spectrum and we construct a non-extremal black hole in the gauged four-dimensional theory. Uplifting this black hole to five dimensions allows us to define a consistent near-horizon limit in which we realize a thermal BTZ geometry; but now with a $\rho$-twisted angular coordinate. To leading order, its entropy matches the Cardy formula of the dual $\rho$-twisted MSW-CFT.
Phrased differently, we conjecture that the $\rho$-twisted MSW-CFT is dual to M-theory on\footnote{Scherk-Schwarz reductions often break supersymmetry spontaneously \cite{Scherk:1978ta}, and the vacuum might be unstable. To deal with this properly, one has to actually start with a $T^6$ instead of a $CY_3$, so that partial supersymmetry can be preserved. This effect is however not relevant for the present leading order calculations.} $AdS^\rho_3 \times S^2 \times CY_3$. The superscript on the $AdS_3$ merely refers to the twisted boundary conditions along the angular coordinate in $AdS_3$.\\

Finally, we present a different example in Section \ref{sec:possdir} based on the D1-D5 system in type IIB on $K3\times S^1$, which is of equal interest as the MSW setup. The line of thought is exactly the same as in the MSW system, except that the R-symmetry is now larger, i.e. $SO(5)_R$ instead of $SU(2)_R$. Furthermore, the CFT is $(4,4)$ instead of $(0,4)$ allowing for $\rho$-twists on either of the chiral sectors. \\

As is evident through this introduction, we will consistently lay emphasis on the two punchlines of this work. One is the microscopic description of a large class of asymptotically flat black holes in R-gauged supergravities. The other is the microscopic realization of the $\rho$-algebras and their conjectured bulk duals.

\section{Rholography}\label{sec:rhoalgebras}

The M5-brane breaks half of the supersymmetry available in M-theory. It carries a chiral $(0,2)$ SCFT in six dimensions. The Lorentz group breaks to $Spin(1,5)$ with an additional $USp(4)$ R-symmetry owing to the transverse directions. 
Its world volume theory consists of 5 scalars $X~\{=X^a, ~ a=1,...,5\}$ (corresponding to the transverse directions of the brane), four six-dimensional Weyl spinors $\psi~\{=\psi_i, ~ i=1,...,4\}$ that obey a symplectic reality condition and an anti-symmetric two form $B_2$ whose field strength is self-dual. Considering M-theory on a $CY_3$ background and placing the M5-brane on a holomorphic compact divisor inside the threefold reduces the symmetry of the world-volume theory. The $USp(4)\simeq SO(5)$ breaks to a $Spin(3) \times Spin(2)$ symmetry\textemdash the $Spin(3)$ comes from the position of the brane in non-compact space while the $Spin(2)$ is owed to the position of the brane in the $CY_3$. Furthermore, the six dimensional local Lorentz group breaks to $Spin(1,1)\times Spin(4)$, and reduces to a $Spin(1,1) \times SU(2) \times U(1)$ symmetry on the M5-string worldsheet \cite{Gaiotto:2006wm}. We may now gauge fix the world-sheet coordinates to align with the target space coordinates to realize the $Spin(1,1)$ Lorentz symmetry on the world-sheet. This is the MSW-CFT and its world-sheet field content can be obtained from the reduction of the M5-brane world-volume fields \cite{Minasian:1999qn}. In this article, we will entirely focus our attention on two symmetries of this field theory\textemdash the $Spin(3)$ that manifests itself as a local $SU(2)$ Kac-Moody algebra in the field theory, and the global $SU(2)$ flavor symmetry. The latter is actually only a symmetry of the algebra, and not necessarily of the CFT. It is the outer automorphism group of the superconformal algebra. It may happen that for large value of the central charge, the outer automorphism group may actually become a symmetry. In the dual bulk, this is the classical supergravity regime. We get back to this point later. Since all the supersymmetry generators are in the right moving sector of the CFT, we can study this $N= 4$ superconformal algebra in its own right. For notational ease, we will call the Kac-Moody gauge group $SU(2)_\eta$ and the outer automorphism group $SU(2)_\rho$. Together, they form the total automorphism group $SO(4)$ of the small $N=4$ superconformal algebra \cite{Schwimmer:1986mf}.\\

It is worth understanding the presence of these symmetries in the different theories of interest. From the black string perspective, the $SU(2)_\eta$ local gauge symmetry is realized as the spherical symmetry of the horizon. It sits inside the Lorentz group of the five-dimensional supergravity theory and similarly, it is also the rotational symmetry goup of a spherical black hole in the four-dimensional supergravity theory. The outer automorphism group, also called the global $SU(2)_\rho$ flavor symmetry when it is a symmetry, however, has roots in the $CY_3$. As we discuss in Section \ref{sec:bulk}, it is the $SU(2)$ R-symmetry of five-dimensional ${\cal N}=2$ supergravity acting on the supersymmetry generators. Upon compactifying on a circle, we will perform the Scherk-Schwarz twist with respect to a $U(1)$ subgroup of this $SU(2)$ R-symmetry. As we will show, in the supergravity regime, this subgroup is actually a symmetry of the action. In the $N=4$ CFT, as has been studied in \cite{Schwimmer:1986mf}, a twisting of the Abelian subgroup of the local $SU(2)_\eta$ is just a gauge symmetry; it can be undone by spectral flow. However, a twisting of the Abelian subgroup of the global $SU(2)_\rho$ symmetry results in an infinite family of $(0,4)$ algebras parametrized by the twisting parameter $\rho$. It cannot be undone because the $U(1)_\rho \subset SU(2)_\rho$ is only an approximate symmetry, in much the same way as the bonus symmetry discussed in \cite{Intriligator:1998ig}. Hence there is no current algebra associated to it, and hence no spectral flow. The twist is `felt' by all the fields in the CFT that transform non-trivially under the $SU(2)_\rho$. This includes, in particular, the supercharges that transform under a doublet representation. Since the twist is under an Abelian subgroup of the R-symmetry, the corresponding five-dimensional supergravity theory realizes it as a specific Scherk-Schwarz reduction on the circle; one that corresponds to an R-gauging in four dimensions. This is a $U(1)$ gauged supergravity theory in four dimensions, and it already indicates that the twist parameter $\rho$ must be related to the $U(1)$ gauge coupling constant. Much like in the un-twisted case, the twisted CFT counts the microstates associated to a black hole in this gauged supergravity theory. 

We will now make the discussion more concrete, by first presenting the small $N=4$ superconformal algebra with $\rho$-twist in Section \ref{N4rhotwist}, and the structure of the holographic dual in Section \ref{rhodual}.

\subsection{Small $N=4$ superconformal algebra with $\rho$-twist}\label{N4rhotwist}

We will call the Virasoro generators $L_m$ (and their corresponding stress tensor $L(z)$), the Kac-Moody generators $T^i$ (with $i = 1, 2, 3$) and the four supercharges $G^{a A}$ (with $a = 1, 2$ and $A = \pm$). Here, $i$ is an $SU(2)_\eta$ triplet index, $a$ an $SU(2)_\eta$ doublet index and $A$ an $SU(2)_\rho$ doublet index. The Operator Product Expansions (OPEs) can be determined from \cite{Schwimmer:1986mf,Yu:1987dh,Defever:1988um}. Dropping the regular terms when $z\rightarrow w$, they are
\begin{align}\label{eqn:OPEs}
L(z) L(w) ~ &= ~ \dfrac{\partial_wL (w)}{z-w} ~ + ~ \dfrac{2 L(w)}{(z-w)^2} ~ + ~ \dfrac{\frac{1}{2} ~ c_R}{(z-w)^4}\ , \nonumber \\
G^{a\pm}(z) G^{b\mp}(w) ~ &= ~ \delta^{ab}~\left(\dfrac{2 L (w)}{z-w} ~ + ~ \dfrac{\frac{2}{3} ~ c_R}{(z-w)^3}\right)
+(\sigma^{i} )^{ab}\left( \dfrac{2~\partial_wT^i(w)}{z-w} ~ + ~ \dfrac{4i~T^i(w)}{(z-w)^2} \right)\ , \nonumber \\
T^i(z) T^j(w) ~ &= ~ i\dfrac{\varepsilon^{ijk}~T^k(w)}{z-w} ~ + ~ \dfrac{\frac{1}{12} ~ c_R~\delta^{ij}}{(z-w)^2}\ , \nonumber \\
L(z) G^{a\pm}(w) ~ &= ~ \dfrac{\partial_wG^{a\pm}(w)}{z-w} ~ + ~ \dfrac{\frac{3}{2}G^{a\pm}(w)}{(z-w)^2}\ , \nonumber \\
L(z) T^i(w) ~ &= ~ \dfrac{\partial_wT^i(w)}{z-w} ~ + ~ \dfrac{T^i(w)}{(z-w)^2}\ , \nonumber \\
T^i(z) G^{a+}(w) ~ &= ~ \dfrac{\frac{1}{2}G^{b+}(w)\left(\sigma^i\right)_{b}{}^{a}}{z-w}\ , \nonumber \\
T^i(z) G^{a-}(w) ~ &= ~ -\dfrac{\frac{1}{2}\left(\sigma^i\right)^{a}{}_{b}G^{b-}(w)}{z-w}\ ,
\end{align}
where $(\sigma^i)^{a}{}_{b}$ are the Pauli matrices.\\ 

As shown in  \cite{Schwimmer:1986mf,Yu:1987dh}, the total automorphism group of these OPEs (and the algebra generated by them) is $SO(4)=SU(2)_\eta\times SU(2)_\rho$. The inner outomorphism group is $SU(2)_\eta$ and corresponds to the current algebra while the outer automorphism group is the global $SU(2)_\rho$. Twists under the Abelian subgroups of the two $SU(2)$ groups are generated by \cite{Schwimmer:1986mf,Yu:1987dh} \textendash 
\begin{align}\label{eqn:twistedsupercharges}
G^{1\pm}(z e^{2 \pi i}) ~ &= ~ e^{\mp i \pi (\rho + \eta)} ~ G^{1\pm}(z)\ , \nonumber \\
G^{2\pm}(z e^{2 \pi i}) ~ &= ~ e^{\mp i \pi (\rho - \eta)} ~ G^{2\pm}(z)\ , \nonumber \\
T^{\pm}(z e^{2 \pi i}) ~ &= ~ e^{\pm 2 \pi i \eta} ~ T^{\pm}(z)\ , 
\end{align}
where $T^{\pm}=T^1\pm iT^2$, while $T^3(z)$ and $L(z)$ are left to be periodic. 
The resulting mode expansion for the supercharges is, therefore, 
\begin{align}
G^{1\pm}(z) ~ &= ~ \sum_{m\in \mathbb{Z}} G^{1\pm}_{m \pm \frac{\rho + \eta}{2} + \frac{1}{2}} ~ z^{-m \mp \frac{\rho + \eta}{2} - 2}\ , \nonumber \\
G^{2\pm}(z) ~ &= ~ \sum_{m\in \mathbb{Z}} G^{2\pm}_{m \pm \frac{\rho - \eta}{2} + \frac{1}{2}} ~ z^{-m \mp \frac{\rho - \eta}{2} - 2}\ .
\end{align}

The usual NS and R sectors have $\rho=0$, with $\eta=0$ and $\eta=1$ respectively. These result in half-integer ($\eta=0$) and integer ($\eta=1$) modes for the supercharges, respectively. For $\rho \neq 0$, one gets inequivalent algebras. In this article, we will exclusively work with non-zero $\rho$. \\ 

Any particular mode can be extracted out of this Laurent series by an appropriate Cauchy integral as 
\begin{align}
G^{1\pm}_{m \pm \frac{\rho + \eta}{2} + \frac{1}{2}} ~ &= ~ \dfrac{1}{2 \pi i}\int \mathrm{d}z ~ z^{m \pm \frac{\rho + \eta}{2} + 1} ~ G^{1\pm}(z)\ , \nonumber \\
G^{2\pm}_{m \pm \frac{\rho - \eta}{2} + \frac{1}{2}} ~ &= ~ \dfrac{1}{2 \pi i}\int \mathrm{d}z ~ z^{m \pm \frac{\rho - \eta}{2} + 1} ~ G^{2\pm}(z)\ .
\end{align}

The anti-commutation relations for the supercharges can now be calculated from this mode expansion and the OPE in \eqref{eqn:OPEs}, using Cauchy's theorem. The result is:\footnote{In this relation, instead of $G^{1-}_{n - \frac{\rho + \eta}{2} + \frac{1}{2}}$, note that we have used a shifted mode $G^{1-}_{n - \frac{\rho + \eta}{2} - \frac{1}{2}}$ such that the latter is the complex conjugate of the generator $G^{1+}_{m + \frac{\rho + \eta}{2} + \frac{1}{2}}$, with $m=-n$. This merely shifts the Laurent expansion appropriately.}
\begin{align}\label{eqn:anti-comm1}
\bigg\{G^{1+}_{m + \frac{\rho + \eta}{2} + \frac{1}{2}} ~ , ~ G^{1-}_{n - \frac{\rho + \eta}{2} - \frac{1}{2}}\bigg\} ~ = ~ &2 ~ L_{m+n} ~ + ~ 2\left(m - n + 1 + (\rho+\eta)\right) ~  T^3_{m+n} ~ \nonumber \\
&+ ~ \dfrac{c_R}{12} ~ \left[\left(2m + 1 + (\rho + \eta)\right)^2 - 1\right] ~ \delta_{m+n,0}\ .
\end{align}
\begin{align}\label{eqn:anti-comm2}
\bigg\{G^{2+}_{m + \frac{\rho - \eta}{2} + \frac{1}{2}} ~ , ~ G^{2-}_{n - \frac{\rho - \eta}{2} - \frac{1}{2}}\bigg\} ~ = ~ &2 ~ L_{m+n} ~ - ~ 2\left(m - n + 1 + (\rho - \eta)\right) ~ T^3_{m+n} ~ \nonumber \\
&+ ~ \dfrac{c_R}{12} ~ \left[\left(2m + 1 + (\rho - \eta)\right)^2 - 1\right] ~ \delta_{m+n,0}\ .
\end{align}

We know that the $\eta$ twist is a gauge redundancy and therefore causes spectral flow. Any physical quantity must be independent of $\eta$. The gauge independent, spectral flow invariant quantities do not depend on $\eta$ and are defined by the relations \cite{Schwimmer:1986mf}
\begin{equation}
L_n \left(\rho,\eta\right) = L_n (\rho) - \eta\, T^3_n(\rho) + \eta^2\, \dfrac{c_R}{12}\,\delta_{n,0}\ ,\quad T^3_n(\rho,\eta)=T_n^3(\rho)-\eta\,\frac{c_R}{6}\delta_{n,0}\ ,\quad T^{\pm}_{n\pm\eta}(\rho,\eta)=T^{\pm}_{n}(\rho)\ ,\nonumber\\
\end{equation}
for the bosonic operators, and
\begin{equation} 
G^{1\pm}_{n\pm \frac{\rho+\eta}{2}+\frac{1}{2}}(\rho,\eta)=G^{1\pm}_{n\pm\frac{\rho}{2}+\frac{1}{2}}(\rho) \qquad G^{2\pm}_{n\pm\frac{\rho-\eta}{2}+\frac{1}{2}}(\rho,\eta)=G^{2\pm}_{n\pm\frac{\rho}{2}+\frac{1}{2}}(\rho)\ ,
\end{equation}
for the modes of the supercharges. Therefore, we see that one way to arrive at the gauge independent quantities from the gauge dependent one, is by setting $\eta = 0$\textemdash this is what we do in the following. The parameter $\rho$ takes values $0\leq \rho \leq 2$, but without loss of generality we can restrict $0\leq \rho \leq 1$ as follows from the periodicity conditions.\\

From the algebra, we can now derive the unitarity constraints on a highest weight state labelled by the eigenvalues $(h,l)$ of $L_0$ and $T_0^3$ respectively. This analysis was done in \cite{Yu:1987dh}, and we state the result here:
\begin{eqnarray}
&&l<\frac{c_R}{12}\ ,\qquad h\geq (1-\rho)\,l+\frac{c_R}{12}\,\rho\,(1-\frac{\rho}{2})\ ,\nonumber\\
&&l=\frac{c_R}{12}\ ,\qquad h =\frac{c_R}{12}\,(1-\frac{\rho^2}{2})\ .
\end{eqnarray}

Since we are interested in black holes with zero angular momentum, we must take $l=0$. The ground state energy then is
\begin{equation}\label{eqn:twistedgroundstate2}
h_0 ~ = ~ \dfrac{c_R}{6} \left( \dfrac{\rho}{2} - \dfrac{\rho^2}{4}\right)\ ,\qquad \left(L_0-h_0\right)|0\rangle=0\ .
\end{equation}

Acting with raising operators in the algebra on this vacuum state, one obtains representations with integer shifts from this ground state. Therefore, a generic state in this sector has a conformal dimension\footnote{Here, $n_L$ and $N_R$ are integers while $h_0$ is a continuous parameter in the space of algebras defined by $\rho$.} $n_R = N_R + h_0$. Therefore, the entropy of the field theory in an excited state with conformal dimensions $n_L$ and $n_R$ in the Cardy regime is given by 
\begin{align}\label{eqn:microentropyinrho}
S_{CFT} &= 2\pi \left( \sqrt{\dfrac{c_L}{6} n_L} ~ + ~ \sqrt{\dfrac{c_R}{6} n_R} \right) \nonumber \\
&= 2\pi \left( \sqrt{\dfrac{c_L}{6} n_L} ~ + ~ \sqrt{\dfrac{c_R}{6} \left( N_R + \dfrac{c_R}{6} \left( \dfrac{\rho}{2} - \dfrac{\rho^2}{4}\right) \right)} \right).
\end{align}
We shall see that this matches with the expectation from the bulk theory, in Section \ref{sec:bulk}, where the momentum along the string is identified with the electric charge of the black hole. Since the field theory is that of an M5-string, the momentum along the string can be calculated to be
\begin{align}\label{eqn:quantizationinrho}
L_0 ~ - ~ \bar{L}_0 ~ &= ~ n_L ~ - ~ n_R \nonumber \\ 
&= ~ n_L ~ - ~ \left(N_R + \dfrac{c_R}{6} \left( \dfrac{\rho}{2} - \dfrac{\rho^2}{4}\right) \right) \ .
\end{align}
This momentum is no longer integer-quantized\textemdash it is shifted by the ground state energy $h_0$. It is worth noticing that the ground state energy vanishes for $\rho=2$, but this value is equivalent to $\rho=0$ as one can see from \eqref{eqn:twistedsupercharges}. The maximum value arises for $\rho=1$, namely $h_0=c/24$. This is precisely the same shift for the ground state energy between the Ramond and Neveu-Schwarz sector. This is not surprising, since $\eta=0$ and $\rho=1$ is equivalent to the Ramond sector which has $\eta=1$ and $\rho=0$.\\

\subsection{$AdS_3^\rho \times S^2$ bulk duals}\label{rhodual}

The twisting of the supercharges in the small $N=4$ superconformal algebra raises the question of what the corresponding operation is in the dual bulk theory that lives on $AdS_3$. A systematic study of the asymptotic dynamics and symmetries of three-dimensional extended supergravity on $AdS_3$ was made in \cite{Henneaux:1999ib} in the Chern-Simons formulation. The $AdS_3$ superalgebra that corresponds to the small $N=4$ superconformal algebra is $SU(1,1|2)/U(1)$ and contains an "inner" $SU(2)$ symmetry that is dual to the $SU(2)_\eta$ current algebra. Furthermore,  it was shown that the twisting of the $SU(2)_\rho$ outer automorphism group corresponds to twisting the periodicity conditions on the gravitini. Indeed, $AdS_3$ has the topology of a disc times a real line, with coordinates $(r,\theta)$ and $t$, and the supergravity fields in three dimensions must be given periodicity conditions in $\theta$ in such a way that the supergravity Lagrangian remains invariant (see Section 6 in \cite{Henneaux:1999ib}). In our notation, following \eqref{eqn:twistedsupercharges} for $\eta=0$, and suppressing the coordinates $r$ and $t$, this means,
\begin{equation}\label{bc-rho}
\psi_\mu^{a\pm}(\theta+2\pi)=e^{\mp i\pi \rho}\psi_\mu^{a\pm}(\theta)\ .
\end{equation}
The three-dimensional gravitini are in general denoted by $\psi_\mu^{aA}$, where the superscripts denote the representation of the R-symmetry in three dimensions. In general, the R-symmetry in $N=4, D=3$ is $SO(4)_R$, but R-symmetry in supergravity is not always a symmetry of the Lagrangian, only of the superalgebra. However, as mentioned in the Introduction and the beginning of this section, our three-dimensional supergravity comes from $\mathcal{N}=2$ in five dimensions, where the R-symmetry is only $SU(2)_R$. The five-dimensional theory is defined on $AdS_3\times S^2$, and after reducing to three dimensions, the R-symmetry enlarges to 
\begin{equation}\label{d=5Rsymm}
D=5: \quad SU(2)_R \qquad \Rightarrow \qquad D=3: \quad SO(4)_R = SU(2)_{\rho} \times SU(2)_{\eta}\ .
\end{equation}
Here, the $SU(2)_\eta$ is now a symmetry and it is gauged, with $SU(2)_\eta$ Chern-Simons gauge fields that are dual to the 
current algebra in the small $N=4$ superconformal algebra. The $SU(2)_\rho$ is the $SU(2)_R$  from five dimensions. It corresponds to the outer automorphism group in the dual CFT. Both these groups ($SU(2)_\rho$ and $SU(2)_R$) are outer automorphisms and are in general not symmetries of the Lagrangian. It is now clear that the index $a=1,2$ denotes the two-dimensional representation of $SU(2)_\eta$ and $A=1,2$ (or in complexified notation $A=+,-$) the one of $SU(2)_\rho$. 
Hence, on the one hand, twisting the periodicity conditions with a $U(1)\subset SU(2)_\rho$ implies twisting a $U(1)\subset SU(2)_R$ in five dimensions. On the other hand, twisting the periodicity conditions on the gravitini with a $U(1)\subset SU(2)_\eta$ can be undone by a gauge transformation or field redefintition in the bulk. In the boundary CFT, this corresponds to spectral flow in the current algebra. \\

While the analysis in \cite{Henneaux:1999ib} was done for pure Chern-Simons supergravity, we assume here that it can be extended to include also matter multiplets and that our reduction from five dimensions can be recasted in this language. This would mean that all fields in five dimensions that have R-charge, will be subject to boundary conditions similar to \eqref{bc-rho}. For hypermultiplets, we discuss this in the next section. \\

Piecing all the above together, we may now conjecture that M-theory on $AdS^\rho_3 \times S^2 \times CY_3$ is dual to the $\rho$ twisted MSW-CFT, which we denote by $(0,4)_\rho$ CFT. By $AdS^\rho_3$, we mean $AdS_3$ with $\rho$-twisted boundary conditions along the angular coordinate in $AdS_3$. This is what we call ``Rholography". Consequently, the $(0,4)_\rho$ theory in an excited state at finite temperature\textemdash as considered above\textemdash accounts for the entropy of a macroscopic excited state above the $AdS^\rho_3$ vacuum. In Section \ref{sec:bulk}, we will show that this excited macroscopic state is precisely a massive, non-extremal $\text{BTZ}^\rho$ black hole, as one might expect; of course, this $\text{BTZ}^\rho$ geometry will also be one with a twisted angular direction. As we will show, in turn, this  $\text{BTZ}^\rho$ geometry appears in the uplift of the four-dimensional black hole using the Scherk-Schwarz mechanism.\\

In closing, let us note that the discussion in this section is rooted in a chiral $N=4$ SCA in two dimensions; therefore, its scope is certainly not limited to just the $(0,4)$ MSW CFT. Let us consider, for instance, the D1-D5 CFT of Strominger and Vafa. It is a $(4,4)$ theory. $\frac{1}{2}$-BPS states in this theory correspond to space-time $\frac{1}{4}$-BPS states. Such $\frac{1}{2}$-BPS states are counted by keeping one of the chiral sectors in the vacuum (using supersymmetry), while exciting the other chiral sector. As was shown in \cite{Strominger:1996sh}, such a count precisely matches the macroscopic entropy of $\frac{1}{4}$-BPS black holes in five-dimensional $\mathcal{N}=4$ supergravity obtained from a Type IIB compactification on a Calabi-Yau twofold times a circle. As was later pointed out in \cite{Horowitz:1996fn}, exciting both chiral sectors of this two dimensional $(4,4)$ theory counts microstates of near-extremal black holes. \\

The reasoning behind rholography works very similarly as for the case discussed before. Compactifications of type IIB on $K3$  yield six-dimensional chiral $(0,2)$ supergravity. The R-symmetry is $SO(5)_R\simeq USp(4)_R\simeq Sp(2)_R$. We then reduce on six-dimensional backgrounds of the type $AdS_3\times S^3$, and in three dimensions with sixteen supercharges, the R-symmetry is in general $SO(8)_R$. The R-symmetry is in general not a symmetry, but since we reduced on $S^3$, an $SO(4)$ subgroup is a symmetry and is gauged. The analogous (to \eqref{d=5Rsymm}) decomposition of the total R-symmetry group is now
\begin{equation}\label{d=6Rsymm}
D=6: \quad SO(5)_R \qquad \Rightarrow \qquad D=3: \quad SO(8)_R \rightarrow SO(4)_{\rho} \times SO(4)_{\eta}\ .
\end{equation}
The $SO(4)_\eta$ is a gauge symmetry and produces two sets of $SU(2)$ current algebras that are present in the left and right-moving sectors of the dual CFT. The $SO(4)_\rho$ further decomposes in two outer automorphism groups of the left and right-moving sectors, and each can be used to give twisted boundary conditions with parameters, say $\rho_L$ and $\rho_R$. As we will argue in Section \ref{sec:possdir}, twisting both sectors would spontaneously break all the supersymmetry of the vacuum in the macroscopic five-dimensional supergravity theory. But the qualitatively new feature arising from considering the ($\rho_L=0$, $\rho_R\neq 0$) D1-D5 system is that the vacuum in the corresponding supergravity theory still breaks supersymmetry spontaneously; but this time, only partially so. Clearly, this results in exactly the same formula \eqref{eqn:microentropyinrho} for the microscopic entropy.\\

While we will move on to the macroscopic discussion corresponding to the MSW CFT in the next section, we will comment on the microscopic counterpart of the D1-D5 CFT in Section 4.\\

\section{Black holes from M-theory and Scherk-Schwarz reductions}\label{sec:bulk}

The four-dimensional black holes we wish to describe in this paper arise from M-theory compactifications on $CY_3\times S^1$. Their microscopic entropy is governed by the MSW (0,4) CFT, and the M5-string is compactified on the $S^1$. As explained in the introduction, we extend the discussion here by imposing a non-trivial Scherk-Schwarz twist along the $S^1$. The twist group element is chosen to be in the $U(1)_R$ subgroup of the $SU(2)_R$ R-symmetry in the five-dimensional supergravity theory. Hence, it acts on the five-dimensional supercharges that transform as a doublet. This way, as we review in the subsection to follow, we generate gauged supergravity in four dimensions with a positive definite scalar potential with a Minkowski vacuum. In the example of this section,  the vacuum spontaneously breaks supersymmetry from ${\cal N}=2$ to ${\cal N}=0$. In our analysis, in this section, we will ignore radiative quantum corrections to the potential and possible worries about instabilities of the vacuum\footnote{There exist Scherk-Schwarz reductions with R-symmetry twists with supersymmetry preserving vacua, as we discuss in the next section. The reader who is too worried about radiative corrections and instabilities of the supersymmetry breaking vacuum mentioned above, might  find the example of the next section more appealing. There, (half of the) supersymmetry in the vacuum is preserved and calculations are under better control. Alternatively, one might start with M-theory on a $T^6$ instead of a $CY_3$, such that partial supersymmetry can remain after the Scherk-Schwarz twist.}. The supersymmetry breaking scale will be proportional to the twist parameter that plays the role of the gauge coupling constant in gauged supergravity. We assume it to be very small, such that quantum corrections are suppressed. Furthermore, we assume the $S^1$ radius $R$ to be much larger than the length scale of the $CY_3$, i.e. $R^6 \gg Vol_{CY_3} \gg l_{11}^6$, where $l_{11}$ is the eleven-dimensional Planck length. In this regime, the supergravity approximation is valid. All particles that carry R-charge in five dimensions (gravitinos, gaugini, and the hypermultiplets) will become massive in four dimensions, with masses set by the supersymmetry breaking scale\textemdash so they will be light. The black holes that we wish to construct are therefore solutions of four-dimensional gauged supergravity, and our set-up allows us to study them in the presence of light charged matter. Since supersymmetry is broken, the only sensible thing to do is to construct non-extremal solutions, though our microscopic matching only works in the near-extremal limit. The uplift of this solution to five dimension is a black string with twisted boundary conditions, and a near horizon geometry that contains a BTZ factor in the near-extremal limit. This near horizon geometry has a holographic dual which is governed by a CFT with a $\rho$-algebra of symmetries as in Section 2, at finite (and small) temperature. \\

For practical purposes, we choose a $CY_3$ with small Hodge numbers, $h_{1,1}=h_{1,2}=1$. Such Calabi-Yau manifolds were constructed in \cite{Braun:2011hd}. As a consequence, the low energy effective action is five-dimensional supergravity coupled to two hypermultiplets and without any vector multiplets.  The Scherk-Schwarz reduction to four dimensions can in this example be carried out in great detail. Nevertheless, we expect our conclusions to hold more generally, for any $CY_3$, and as a result for more general hypermultiplet couplings. We therefore start Section 3.1 with some general statements about Scherk-Schwarz reductions in supergravity, and then specify our model in more detail. In Section 3.2, we discuss black hole solutions while in Section 3.3, we uplift them to five dimensions and argue for a match of their macroscopic entropy with the Cardy-formula \eqref{eqn:microentropyinrho}. \\

\subsection{R-Symmetry and Scherk-Schwarz reduction}

A generic compactification of M-theory on a $CY_3$ yields an effective five-dimensional theory of $\mc N=2$ supergravity coupled to $h_{1,1}-1$ vector multiplets and $h_{1,2}+1$ hypermultiplets \cite{Cadavid:1995bk}. Further compactification on a circle $S^1$ gives an additional Kaluza-Klein vector multiplet (so $h_{1,1}$ in total) and the same number of hypermultiplets as in five dimensions. 
The effect of doing a Scherk-Schwarz twist on $S^1$ is to yield four-dimensional {\it gauged} $\mc N=2$ supergravity with a gauge group $U(1)$. 
Our setup follows the treatment and the analysis of \cite{Looyestijn:2010pb,Andrianopoli:2004im}, and we use the conventions of \cite{Looyestijn:2010pb}. The five-dimensional metric is decomposed as
\begin{equation}\label{KK}
{\rm d}s^2_{(5)}=R^{-1}{\rm d}s^2_{(4)}+R^2({\rm d}z+A^0)^2\ ,
\end{equation}
where $z\sim z+2\pi $ is the coordinate along the circle, and $R$ denotes the radius of the circle above a base point $x$. All length scales are measured in terms of the eleven-dimensional Planck units.  Finally, $A^0$ is the Kaluza-Klein vector that we also call the four-dimensional graviphoton. Five-dimensional gauge fields decompose as\footnote{For a supergravity theory obtained as a compactification of M-theory on a $CY_3$ with Hodge numbers $h_{1,1}$ and $h_{1,2}$, the indices in \eqref{Vg} are $\Lambda,\Sigma\in\{0,1,...n_v\}$ and $I,J\in\{1,...n_v\}$, with $n_v$ the number of vector multiplets, and  $u,v\in\{1,...,4n_h\}$,  with $n_h=h_{1,2}+1$ the number of hypermultiplets. In five dimensions, we have $n_v=h_{1,1}-1$ and in four dimensions, we have $n_v=h_{1,1}$. The number of hypermultiplets stays the same in five and four dimensions.}
\begin{equation} \label{eq:vectorKK}
A^{I}_{(5)}=A^{I}_{(4)}+a^{I}({\rm d}z-A^{0})\ ,
\end{equation}
with $a^{I}$ four-dimensional scalars. They combine into complex scalars with the real scalars $h^{I}$ of the five-dimensional vector multiplet
\begin{equation}
t^{I}=a^{I}-i R\,h^{I}\ .
\end{equation}
All these fields have zero $SU(2)_R$ R-charge in five dimensions, so they reduce to four dimensions just like in a Kaluza-Klein reduction. Their zero modes are massless. The non-trivial Scherk-Schwarz twist here is performed only on those quantities that transform under the R-symmetry. These include the supercharges, hence the fermions, and the hypermultiplet scalars. These fields get a non-trivial $z$-dependence, different from a Kaluza-Klein expansion of a periodic field. As a consequence, what used to be the massless zero modes in a Kaluza-Klein scheme, now become massive modes, with masses proportional to the twist parameter. These modes are taken to be very light compared to the higher Kaluza-Klein modes. This can be achieved by taking the twist parameter to be small. To be more concrete, we can define the Scherk-Schwarz twist on the supercharges $Q^{A};~A=1,2$, which form a doublet under $SU(2)_R$, as
\begin{eqnarray}\label{bcQ}
Q^A(x^\mu, z+2\pi)&=&\left(e^{2i\pi \alpha\sigma_{3}}\right)^{A}{}_B\,Q^B(x^\mu,z)\ ,
\end{eqnarray}
for a Scherk-Schwarz phase $\alpha$ belonging to the $U(1)_R\subset SU(2)_R$, and with $\sigma_3$ being the third Pauli matrix. A similar transformation holds for the gravitini $\psi_\mu^{A}$ and for the gaugini $\lambda^{AI}$. Comparing with the twist on the worldsheet supercharges in \eqref{eqn:twistedsupercharges} with $\eta=0$, we identify
\begin{equation}
\alpha=\frac{\rho}{2}\ .
\end{equation}
The justification for this was given before, namely that we identify the bulk $SU(2)_R$ symmetry with the worldsheet $SU(2)_\rho$ outer automorphism group. This is because the $S^1$ we twist on, is the same as the $S^1$ we wrap the M5-string around. Similarly, the $S^1$ we twist on is the same $S^1$ that becomes part of the $AdS_3$ in the near horizon geometry of the black string. In essence, the coordinate $z$ is equal to $\theta$ used in \eqref{bc-rho}. So the periodicity conditions we used on the supercharges \eqref{bcQ} and gravitini are also the same as in \eqref{bc-rho}. \\

Any (complex) field $\Phi(x,z)$ with twisted periodicity conditions has a mode expansion
\begin{equation}
\Phi(x,z)=e^{i\alpha z}\sum_{n=-\infty}^{+\infty}\Phi_n(x)e^{inz}\ .
\end{equation}
In a Scherk-Schwarz reduction, we restrict to the $n=0$ mode in the expansion. In other words, we give the five-dimensional field a particular $z$-dependence that satisfies
\begin{equation}
\Phi(x,z)=e^{i\alpha z}\Phi_0(x) \qquad \Longrightarrow \qquad \partial_z \Phi = i\alpha \Phi\ .
\end{equation}
The effect of this is that the four-dimensional field becomes both charged and massive, with $m^2=q^2$ in the appropriate units. The masses will be proportional to $\alpha$ and inversely proportional to the radius $R$, and we give an explicit example at the end of this subsection.\\

Applied to the case at hand, we get
\begin{eqnarray}
\p_z Q^A={i \alpha} \sigma_{3\phantom{A}B}^{\phantom xA} Q^B\ ,
\qquad
\p_z\psi^A_\mu={i \alpha}\sigma_{3\phantom{A}B}^{\phantom xA}\psi^B_\mu\ ,
\qquad
\p_z\lambda^{AI}={i \alpha}\sigma_{3\phantom{A}B}^{\phantom xA}\lambda^{BI}\ .
\end{eqnarray}
These fermionic fields transform with the same Scherk-Schwarz phase, because they are in the same (doublet) representation of the $SU(2)_R$ symmetry. \\

In the hypermultiplet sector, both scalars and fermions transform under this twist. The scalars parametrize a quaternion-Kähler manifold of dimension $4n_h$, with metric $h_{uv}$, and the holonomy group is contained in $SU(2)_R\times USp(2n_h)$. For a given hypermultiplet scalar manifold which is a coset of the form $G/H$, the maximal compact subgroup always contains an $SU(2)_R$ factor. So, homogeneous quaternion-Kähler manifolds always contain $SU(2)_R$  isometries, and hence the Scherk-Schwarz twist can be implemented using the $U(1)_R\subset SU(2)_R$ Killing vector (we add a subscript ``0" to the Killing vector for later notational purposes), 
\begin{equation}\label{SS-q}
\p_z q^{u}=\alpha k_0^{u}(q)\ ,
\end{equation}
so the Scherk-Schwarz twist is in general non-linearly realized on the real hypermultiplet scalars. One can write down a similar formula for the hyperini, using the results of \cite{deWit:1998zg,deWit:1999fp}. Since this is not very insightful, we refrain from giving explicit expressions here. \\

In general, Scherk-Schwarz twists lead to gauged supergravities in one dimension lower, with supersymmetry preserved at the level of the action. 
Gauged supergravities have scalar potentials $V_g$ which are positive-definite for Scherk-Schwarz reductions. Furthermore, they typically allow Minkowski vacua with spontaneously broken supersymmetry. The original references on the topic are \cite{Scherk:1978ta,Scherk:1979zr}. Some other useful literature can be found in e.g. \cite{Bergshoeff:1997mg,Dabholkar:2002sy}.

In four-dimensional $N=2$ supergravity, the bosonic sector of the theory is generically (for electric gaugings) described by the action \cite{Andrianopoli:1996cm} 
\begin{eqnarray}\label{S4dSS}
S_{4d}&=&\int\,
\frac R2 *\mathbf{1}-\gamma_{i\bar{k}}dt^i\wedge *d\bar{t}^{\bar{k}}
+\frac{1}{4}\mathcal I_{\Lambda\Sigma} F^\Lambda\wedge*F^{\Sigma}+\frac{1}{4} \mathcal R_{\Lambda\Sigma}F^{\Lambda}\wedge F^{\Sigma}+\nn\\
&&\hspace{10pt}+h_{uv}Dq^u\wedge * Dq^v-V_g *\mathbf{1}\ ,
\end{eqnarray}
and the potential has a universal form for generic gaugings described by  \cite{Andrianopoli:1996cm}\footnote{We use the conventions of \cite{Looyestijn:2010pb}, which differ from  \cite{Andrianopoli:1996cm} by factors of two in the potential and gauge kinetic terms. One can switch between the conventions by rescaling our four-dimensional metric $g\rightarrow \frac{1}{2}g$ and then multiplying the action by 2. This has the effect of rescaling our potential with an overall factor of $\frac{1}{2}$ and our gauge kinetic terms with an overall factor of 2, while the scalar kinetic terms and Einstein-Hilbert term, normalized as ${\cal L}=\frac{1}{2}{\sqrt{-g}}R(g)$, remain the same.}
\begin{eqnarray}\label{Vg}
V_g&=&2g^2\left(\gamma_{i\bar k}k^i_\Lambda k^{\bar k}_{\Sigma}+4h_{uv}k^u_\Lambda k^v_{\Sigma}\right)
\bar L^\Lambda L^\Sigma+2g^2\left(U^{\Lambda \Sigma}-3 \bar L^\Lambda L^\Sigma\right)
P^x_\Lambda P^x_\Sigma \ .
\end{eqnarray}
where $i,j=1,..,n_v$, $u,v=1,..,4n_h$, $\Lambda,\Sigma=1,..,n_v+1$.
In this formula, $g$ is the gauge coupling, $k^i_\Lambda$ and $k^u_\Lambda$ are Killing vectors of the special K\"ahler and quaternionic isometries respectively, $\gamma_{ i\bar k}$ is the metric of the special K\"ahler manifold with holomorphic coordinates $t^i$, and $h_{uv}$ the metric of the quaternionic manifold with coordinates $q^u$. Notice that the hypermultiplet scalars now appear with a covariant derivative $D_\mu q^u=\partial_\mu q^{u}+k^{u}_\Lambda A_\mu^{\Lambda}$, since they are charged under the Kaluza-Klein field, as discussed above. The symplectic sections $L^\Lambda$ are defined from the holomorphic ones by $L^\Lambda=e^{K/2}X^\Lambda$, where $K$ is the Kähler potential, and we use special coordinates such that $X^0=1$. For more on conventions and properties on special geometry, see \cite{Andrianopoli:1996cm}.  $P^x_\Lambda; x=1,2,3$ are the moment maps that can be computed from the quaternionic Killing vectors. Finally, $U^{\Lambda\Sigma}$ is the symmetric tensor defined on any special Kähler manifold. The precise definition is not important here, since the last term in \eqref{Vg} will vanish in our case. \\

In our setup, only the Kaluza-Klein vector $A^{0}$ from \eqref{KK} is involved in the gauging, and this gauge field is labeled by indices $\Lambda, \Sigma =0$. Even if other gauge fields are present, they do not take part in the gauging in the sense that no fields are charged under them.
The only relevant moment map is therefore $P^x_0$, and thus the only relevant Killing vector of quaternionic isometries is $k^u_0$, which we specify below. Moreover, by properties of special geometry it holds that $(U^{0 0}-3 \bar L^0 L^0)\equiv0$ in the large radius limit,  so the last terms in the potential \eqref{Vg} vanish. \\

Since we will perform a Scherk-Schwarz twist with respect to the R-symmetry, and the scalars in the vector multiplet have no R-charge, the corresponding four-dimensional spectrum should have scalars in the vector multiplets that remain massless and uncharged. This is simply achieved by choosing the gauging of a compact $U(1)$ isometry in the hypermultiplet scalar manifold only,  thus implying {$k^i_\Lambda=0$ for every $\Lambda=0,1,..,n_v$. The potential we consider in this work is then of the no-scale form
\begin{eqnarray}\label{our-Vg}
V_g&=&2g^2\left(4h_{uv}k^u_0 k^v_{0}\right)
\bar L^0 L^0=8g^2h_{uv}k^u_0 k^v_{0}e^K\ ,
\end{eqnarray}
and is positive definite. Using the relations, in the conventions of \cite{Looyestijn:2010pb},
\begin{equation}
e^{-K}=8R^3\ ,\qquad {\sqrt {-g_{(5)}}}=\frac{\sqrt {-g_{(4)}}}{R}\ ,
\end{equation}
one can interpret this as a potential coming from the dimensional reduction of the hypermultiplet scalars' kinetic terms \cite{Andrianopoli:2004im}. Indeed, using \eqref{SS-q}, we find
\begin{equation}\label{SS-pot}
{\sqrt {-g_{(5)}}}h_{uv}\partial_z q^{u} \partial_z q^{v} g^{zz}=\sqrt {-g_{(4)}}\,V_g = \sqrt {-g_{(4)}}\,\frac{g^2}{R^3}h_{uv}k^u_0 k^v_{0}\ .
\end{equation}

From this, one can see two possible types of vacua, both of which are Minkowski. The first one is to have the Killing vectors finite and non-zero in the vacuum; the potential is then of the runaway type and the theory decompactifies. We are not considering this option since in our case, the Killing vectors of the R-symmetry will have fixed points and vanish in the vacuum. $R$ is then a flat direction, and the potential is called no-scale. Therefore, we can freely take the radius to be large, such that $R^6\gg Vol_{CY_3}$. \\ 

The masses of the particles in the spectrum follow from expanding fluctuations around the vacuum to quadratic order, and involve the derivatives of the Killing vectors which need not vanish in the vacuum. We refer to \cite{Andrianopoli:1996cm} for general expressions of the mass matrices. Furthermore, the Scherk-Schwarz reduction also generates terms proportional to the Kaluza-Klein vector $A^0$, from which one can determine that the charge\footnote{In computing the charge, one must take care of the correct normalization of the Kaluza-Klein vector. In the conventions of \cite{Looyestijn:2010pb}, 
the kinetic term for $A^0$ is ${\cal L}=-\frac{R^3}{8}F_{\mu\nu}F^{\mu\nu}$, so one needs to rescale the gauge fields $A^0\rightarrow \frac{\sqrt 2}{R^{3/2}} A^0$ to have a canonically normalized Maxwell field.} is equal to the mass, $m^2=q^2$. 
Finally, for \eqref{SS-pot} to hold, we identify the Scherk-Schwarz twist parameter with the gauge coupling constant
\begin{equation}\label{g-rho}
\alpha=g\qquad \Longrightarrow \qquad g=\frac{\rho}{2}\ .
\end{equation}

It is important to notice that in the vacuum, the bosonic part of the Lagrangian becomes that of ungauged supergravity. Indeed, in the vacuum, the potential vanishes and all covariant derivatives on the hypermultiplet scalars become ordinary ones since the covariant derivatives involve Killing vectors that vanish in the vacuum. The hypers can therefore be frozen to their vevs. The scalars in the vector multiplets remain neutral. As a consequence, any bosonic solution of the equations of motion in ungauged supergravity without hypermultiplets can be imported into the R-gauged supergravity theory. This observation will be important when we discuss black hole solutions in Section 3.3.\\

\section*{Example}

The derivation of the scalar potential in the four-dimensional theory holds for any choice of  $U(1)$ Scherk-Schwarz gauging from five to four dimensions, gauged by the graviphoton (Kaluza-Klein vector $A^{0}$), for a generic $CY_3$-compactification. To exemplify our strategy further, we now choose a particular model, namely the case in which the $CY_3$ has $h_{1,1}=h_{1,2}=1$, as discussed at the beginning of this section. Such a compactification gives a five-dimensional $\mc N=2$ supergravity theory with no vector-multiplets and $n_h=2$ hypermultiplets whose scalar manifold is the c-map of $SU(1,1)/U(1)$. This has been extensively studied in \cite{Bodner:1989cg} and \cite{Ferrara:1989ik}, for example. A result of these studies is that the quaternionic manifold is $G_{2(2)}/SO(4)$ where $SO(4)=SU(2)_R\times SU(2)$. 
We parametrize it by introducing coordinates 
\begin{equation} \label{eq:G2SO4Coords}
q^u=(\phi,\varphi,\chi,a,\xi^0,\xi^1,\tilde\xi_0,\tilde\xi_1)\ .
\end{equation}
Here, $\varphi$ and $\chi$ form a complex structure modulus, the $\xi$ and $\tilde\xi$ come from the periods of three-form in eleven dimensions restricted to the $CY_3$, and $a$ is the dual of the three-form, restricted to five dimensions. Finally, the (dimensionless) volume-modulus of the $CY_3$\textemdash measured in terms of eleven-dimensional Planck units\textemdash is given by
\begin{equation}
Vol_{CY_3}=e^{-2\phi}\ .
\end{equation}
In these coordinates, the metric is
\begin{eqnarray} \label{cmapmetricG2SO4}
h_{uv} dq^u dq^v = &&\ d\phi^2 + 3(d\varphi)^2 + \tfrac{3}{4} e^{4\varphi} (d\chi)^2  + \tfrac{1}{4} e^{4\phi} \left[ da +  \xi^0 d\tilde{\xi}_0 + \xi^1 d\tilde{\xi}_1 - \tilde{\xi}_0 d\xi^0 - \tilde{\xi}_1 d\xi^1  \right]^2 \nonumber \\
&& \ +\tfrac{1}{2} e^{2\phi-6\varphi} (d\xi^0)^2
+\tfrac{1}{2} e^{2\phi-2\varphi} \left[d\xi^1 -\sqrt{3} \chi d\xi^0 \right]^2 \nonumber \\
&& \ +\tfrac{1}{2} e^{2\phi+2\varphi} \left[ d\tilde{\xi}_1 - \sqrt{3} \chi^2 d\xi^0 + 2\chi d\xi^1  \right]^2 \nonumber \\
&& \ +\tfrac{1}{2} e^{2\phi+6\varphi} \left[ d\tilde{\xi}_0 + \sqrt{3} \chi d\tilde{\xi}_1 -\chi^3 d\xi^0 + \sqrt{3} \chi^2 d\xi^1  \right]^2 \, ; 
\end{eqnarray}
see equation (5.4) of \cite{Cassani:2012pj} for a similar parametrisation. \\

The scalar potential can be found from the Killing vector belonging to the $U(1)_R\subset SU(2)_R\subset SO(4)$ isometry. The explicit form for this Killing vector is given in the appendix, using the parametrization (\ref{cmapmetricG2SO4}) for the metric on the coset $G_{2(2)}/SO(4)$. 
A Minkowski vacuum is then obtained for the values of the fields which are a vanishing locus for the Killing vector $k^u_0=0$, and thus, for the choice  \eqref{KillR} in appendix \ref{app:KVs}, 
\begin{eqnarray} \label{VanishingPoint}
\chi=a=\xi^0=\xi^1=\tilde \xi_0=\tilde \xi_1=0 \; , \quad e^{4\phi} = \gamma^2 \delta^4 \; , \quad e^{4\varphi} = 3 \gamma^2 \ .
\end{eqnarray}
The parameters $\gamma$ and $\delta$ specify the choice of the Killing vector as seen inside $G_{2(2)} $\textemdash this may be seen from \eqref{eqn:G2killings1} and \eqref{eqn:G2killings2}. The volume, therefore, may be chosen to be large by specifying an appropriate Killing vector with large $\delta$, for example.\\

Expanding around the vacuum, one can determine the masses of the hypermultiplet scalars. In four-dimensional Planck units\footnote{The scalar potential in \eqref{our-Vg} contains a $\kappa_4^{-2}$, so all masses scale with the four-dimensional Planck mass in our model. The gauge coupling constant $g$ is dimensionless.} , they are found to be
\begin{eqnarray}
m^2_{(0)}=\frac{g^2}{R^3}\ ,
\end{eqnarray}
and are fully degenerate, i.e. all eight hyperscalars have the same mass. In the fermionic sector, all the fields are charged under $U(1)_R\subset SU(2)_R$. The gravitini undergo a super-Higgs mechanism and become massive by eating up the gaugini. Their mass eigenvalues can be computed from the moment maps (see e.g. \cite{Andrianopoli:1996cm} for more details on the gravitino mass matrix). In four-dimensional Planck units, we again find 
\begin{eqnarray}
m^2_{(3/2)}&=&
\frac{g^2}{R^{3}}\ .
\end{eqnarray}
The gravitini mass sets the supersymmetry breaking scale. It is very small in the regime we are working in, namely large radius $R$ and small coupling $g$. This provides an argument why radiative corrections might be suppressed. \\

The fermionic sector in the hypermultiplets contains two Dirac spinors, one for each hypermultiplet. Equivalently, there are four chiral components $\zeta_\alpha; \alpha=1,...,4$. Their Dirac masses are found to be 
\begin{eqnarray}
m_{(1/2)}=0\ ,\qquad {\rm and} \qquad m_{(1/2)}=m_{(3/2)}\ .
\end{eqnarray}
The chiral components in each hypermultiplet then have the same masses, but with double degeneracy. \\

\subsection{Black holes in R-gauged supergravity}

The example and general considerations in the previous subsection illustrate the following: After freezing the hypermultiplets to their expectation values, the Killing vectors vanish and so does the scalar potential. Turning to the bosonic sector described by the action \eqref{S4dSS}, the resulting supergravity Lagrangian after freezing the hypers \eqref{VanishingPoint} is precisely that of ungauged supergravity. The covariant derivatives on the hypermultiplet scalars become ordinary derivatives, and so the hypermultiplets decouple classically. We have already mentioned that the scalar of the vector multiplet  is a flat direction in the Minkowski vacuum obtained by Scherk-Schwarz twist on the $U(1)_R$ isometry. In particular, the vector multiplet equations of motion decouple from the hypermultiplet ones, and are totally insensitive to the Scherk-Schwarz twist: they are effectively the same as the equations of ungauged supergravity. Therefore, every solution of the ungauged supergravity bosonic Lagrangian for the metric and the scalars of the vector multiplets is also automatically a solution of the Scherk-Schwarz reduced theory around the Minkowski vacuum where the hypermultiplets are stabilized. This has been discussed in the context of near-horizon supersymmetry already in \cite{Hristov:2012nu} and more recently in the context of near-horizon dimensional reduction in \cite{Hristov:2014eba}. Constructions of black hole solutions in gauged supergravities with maximal supersymmetry preserving vacua can be found in \cite{Hristov:2010eu}. The black holes we consider here live in supersymmetry breaking vacua. The energy scale set by the temperature of the black hole is supposed to be larger than the supersymmetry breaking scale, yet still low enough such that the specific heat remains positive. For black hole temperatures lower than the supersymmetry breaking scale, the massive modes first need to be integrated out.\\

We then consider a non-extremal black hole\textemdash a solution of the theory \eqref{S4dSS} around the Minkowski vacuum coupled to $n_v=1$ vector multiplet. This corresponds to the dimensional reduction of five-dimensional minimally coupled supergravity. We further truncate to zero axions  and consider the case of one electric charge $q_0$, and one magnetic charge, $p^1$, with the scalar field $t$ being the coordinate of $SU(1,1)/U(1)$. The black hole is a solution of the Einstein, scalar and Maxwell equations\footnote{Here we switch back to the conventions used in  \cite{Andrianopoli:1996cm} that are most commonly used in the black hole literature.}
\begin{eqnarray}
&&R_{\mu\nu}-\frac R2 g_{\mu\nu}=g_{\mu \nu }\left(-\gamma_{t\bar t}\partial_\mu t\partial^\mu \bar t+
\frac14 \mc I_{\Lambda \Sigma}F_{\mu \nu}^\Lambda F^{\Sigma\,\mu\nu}\right)- \mc I_{\Lambda \Sigma}F_{ \alpha \mu}^\Lambda F^{\Sigma\, \alpha}_{\phantom{\Sigma \alpha}\nu}+2 \gamma_{t\bar t} \partial_\mu t\partial_\nu \bar t\ ,\nn\\
&&-\frac1{\sqrt{-g}}\partial_\mu\left(\sqrt{-g}\partial^\mu \bar t\right)-\gamma^{t\bar t}\partial_{\bar t}\gamma_{t\bar t}\,\partial_\mu \bar t \partial^\mu t=\gamma^{t\bar t}\partial_t(\mc I_{\Lambda \Sigma})F_{\mu \nu}^{\Lambda}F^{\Sigma\, \mu\nu}\ ,
\qquad \gamma_{t\bar t }=\frac{3}{4\im(t)^2}=(\gamma^{t\bar t})^{-1}\ ,\nn\\
&&\partial_\mu\left(
\sqrt{-g}\mc I_{\Lambda \Sigma} F^{\Lambda\,\mu\nu}\right)=0\ ,
\end{eqnarray}
where there is no summation on the $t$ and $\bar t$ indices since the scalar manifold is of complex dimension $1$.  The setup of this solution corresponds to a particular case of \cite{Galli:2011fq}. \\

The metric of the black hole solution in the region outside the horizon is given by 
\begin{eqnarray}\label{non-extr-r-coord}
{\rm d}s^2_{(4)} &=&
-e^{2U(r)} {\rm d}t^{2} + e^{-2U(r)}  {\rm d}r^{2} 
+e^{-2U(r)} f(r) {\rm d}\Omega^{2}_{(2)}\, ,
\end{eqnarray}
with $f(r)=(r-r_+)(r-r_-)$ and ${\rm d}\Omega_{(2)}^2={\rm d}\theta^2+\sin\theta^2 {\rm d}\phi^2$. We have denoted the inner and outer black hole horizons by $r_\pm=r_*\pm r_0$, while $r_*^2=2\sqrt{|q_0(p^1)^3|}$ refers to the radius of the extremal solution obtained by taking the limit $r_0\rightarrow0$. The warp factor $U(r)$ and the purely imaginary scalar field\textemdash parametrized as $t(r)=-i \lambda(r)$\textemdash are determined in terms of two harmonic functions as
\begin{eqnarray} \label{defFuncs}
e^{-2U(r)}=\frac{r-r_-}{r-r_+}4\sqrt{\mc I_0(\mc I_1)^2}\ ,\qquad 
\lambda(r)=\sqrt{\frac{\mc I_0}{\mc I_1}}\ ,
\end{eqnarray}
where 
\begin{eqnarray}
{\mathcal{I}}_{0} &=&\frac{R^{3/2}}{2}\,\,\frac{r-r_*+r_0\sqrt{1+\frac{2q_0^2}{R^3r_0^2}}}{r-r_*+r_0}\ ,
\qquad
{\mathcal{I}}_{1}=\frac{1}{2\sqrt{ R}}\,\,\frac{r-r_*+r_0\sqrt{1+\frac{2(p^1)^2R}{r_0^2}}}{r-r_*+r_0}\ .
\end{eqnarray}

We note that the scalar field at infinity becomes the dilaton of the Minkowski vacuum discussed in the previous section, $R$, which is a free parameter. The gauge fields of the theory are\textemdash with  $F^{\Lambda}=\frac12F^\Lambda_{\mu\nu}{\rm d}x^\mu \wedge {\rm d}x^\nu$\textemdash 
\begin{eqnarray} \label{eq:GaugeFieldsBH}
F^0&=&\frac{q_0}{ R^3}\frac{1}{\left(r-r_*+r_0\sqrt{1+\frac{2q_0^2}{R^3r_0^2}}\right)^{\hspace{-4pt}2}} \ dt \wedge dr\ ,
\qquad
F^1=p^1\sin\theta d\theta\wedge d\phi\ .
\end{eqnarray}
One then finds the entropies associated to the inner and outer horizons to be
\begin{eqnarray}
\frac{S_\pm}{\pi}&=&
\left(
r_{0} \pm \sqrt{r^{2}_{0} + \frac{2q_0^2}{R^3}} 
\right)^{1/2}
\left(
r_{0} \pm \sqrt{r^{2}_{0} + 2(p^1)^2 R} 
\right)^{3/2}
\, .
\end{eqnarray}
The non-extremal parameter is related to the thermodynamic quantities of the black hole by $r_0=2S_+T$, with the temperature being $T=\frac{\kappa}{2\pi}=\frac{r_+-r_-}{4\pi r_+^2}$.
In the extremal case, $r_0=0$, the radius $R$ drops out of the entropy formula and we obtain the well-known result 
\begin{equation}
S=2\pi{\sqrt {q_0(p^1)^3}}\ ,
\end{equation}
that has been reproduced microscopically for BPS black holes in ungauged supergravity. The mass of the non-extremal black hole is
\begin{eqnarray} \label{MassBH}
M = 
\tfrac{1}{4}\left[3\sqrt{r^{2}_{0}+2(p^1)^2 R} 
+\sqrt{r^{2}_{0}+ \frac{2q_0^2}{R^3}}\right]\, .
\end{eqnarray}
This solution has a smooth $T\rightarrow 0$ limit but, in the absence of supersymmetry, its stability is no longer guaranteed. Therefore, on physical grounds, we choose to work with a non-extremal black hole. \\

\subsection{Uplift to 5 dimensions}

Turning the circle reduction to the four-dimensional theory (\ref{S4dSS}) around, the 4D black hole \eqref{non-extr-r-coord}\textemdash\eqref{eq:GaugeFieldsBH} can be uplifted to a five-dimensional black string. We will now demonstrate that, close to extremality, the near horizon region of this black string displays a  BTZ factor.  In this 5D near-horizon region, the scalar $\lambda(r)$ supporting the back string becomes independent of the radial variable $r$. For simplicity of presentation, we set the scalar to constant already in four-dimensions before uplifting. We have verified that this gives the same result as uplifting the full black hole solution  \eqref{non-extr-r-coord}\textemdash\eqref{eq:GaugeFieldsBH} and then taking the scalars to be constant, since every correction to the near-horizon physics from the running scalars starts at higher orders. \\

We thus set the scalar $\lambda$ in \eqref{defFuncs} to its attractor value,  $\lambda= R=\sqrt{\frac{q_0}{p^1}}$, everywhere. It seems that this choice fixes the dilaton of the Minkowski vacuum; however, one must remember that the constant scalars case is just a shortcut to identify the 5d near horizon region. So, in this case, fixing the value of $R$ has no physical meaning and one should simply treat this as a calculational trick. Trading the non-extremality parameter $r_0$ for the mass $M$ through (\ref{MassBH}) and changing to a new radial variable
\begin{eqnarray}
\tilde r=r-r_*+M\ ,
\end{eqnarray}
the solution \eqref{non-extr-r-coord}\textemdash\eqref{eq:GaugeFieldsBH} becomes Reissner-Nordstr\"{o}m,
\begin{eqnarray} \label{eq:RN4d}
{\rm d}s^2_{(4)} = - \left( 1-\tfrac{2M}{\tilde r}+\tfrac{r_*^2}{\tilde r^2} \right) {\rm d}t^{2} + \frac{ {\rm d}\tilde r^{2}}{ 1-\frac{2M}{\tilde r}+\frac{r_*^2}{\tilde r^2} } 
+\tilde r^2 {\rm d}\Omega^{2}_{(2)} \; , \quad 
F = \frac{2r_*}{\tilde r^2} \, {\rm d}t \wedge {\rm d}\tilde r \; ,
\end{eqnarray}
where $F \equiv \lambda^{3/2} F^ 0 = \frac{1}{\sqrt{3}} \lambda^{1/2} *F^1$. Using the formulae \eqref{KK} and \eqref{eq:vectorKK} for a single vector multiplet, this solution uplifts on the circle parametrised by the angle $z$ to the five-dimensional black string
\begin{eqnarray} \label{5Dstring}
&& {\rm d}s^2_{(5)}  = \sqrt{\tfrac{p^1 }{q_0}} \Big( - \big( 1-\tfrac{2M}{\tilde r} + \tfrac{r_*^2}{\tilde{r}^2} \big) {\rm d}t^2 + \frac{{\rm d} \tilde{r}^2}{  1-\tfrac{2M}{\tilde r} + \tfrac{r_*^2}{\tilde{r}^2}  }  + \tilde{r}^2 {\rm d}\Omega_{(2)}^2  \Big) + \tfrac{q_0 }{p^1 } \big( {\rm d}z + \sqrt{\tfrac{2(p^1)^3 }{q_0 }} \tfrac{1}{\tilde{r}} \ {\rm d}t \big)^2 \; ,  \nonumber \\
&& F_{(5)}  = \sqrt{6} p^1 \,  \sin \theta \, {\rm d} \theta \wedge  {\rm d}\phi \; ,
\end{eqnarray}
where $F_{(5)}$ is the field strength of the gauge field in  \eqref{eq:vectorKK}. \\

Let us now exhibit how the announced BTZ factor arises from the solution \eqref{5Dstring} in the near horizon region, close to extremality. To this end, we rescale
\begin{eqnarray}
\tilde r&\rar& r_*+ \epsilon \rho\ ,\qquad M\rar r_*+\epsilon^2 \frac{\rho_0^2}{2r_*}\ ,\qquad
t\rar \frac{1}{\epsilon}r_*^2\tau\ ,\qquad z \rar \left(\frac{p^1}{q_0}\right)^{3/4}\big(r_*\, \varphi- t\big)\qquad
\end{eqnarray}
following e.g. \cite{Chen:2010bsa}  and then let $\epsilon\rar 0$. In this near horizon, near extremal limit, the five-dimensional metric \eqref{5Dstring} becomes
\begin{eqnarray}\label{metBTZuplift}
 {\rm d}s^2_{(5)}  &=& 2(p^1)^2
\left(
 - (\rho^2-\rho_0^2){\rm d} \tau^2 + 
\frac{{\rm d} \rho^2}{ (\rho^2-\rho_0^2) }  +
\left( {\rm d} \varphi - \rho \ {\rm d}\tau \right)^2
 \right) +2(p^1)^2 \, {\rm d}\Omega_{(2)}^2 \; , 
\end{eqnarray}
which is the direct product of a BTZ metric and a two-sphere $S^2$, of radius $2(p^1)^2$. To see this more explicitly, we identify $\rho_0 = \frac{r_+ ~ + ~ r_-}{2 \ell}$ and make a further change of coordinates 
\begin{eqnarray} \label{changeBTZ}
\rho = \frac{1}{\ell \left(r_+ - r_-\right)} \left( r^2 -\frac{1}{2}(r_+^2 + r_-^2) \right) , \quad 
\tau = 2 \left(- \frac{t}{\ell} +  \phi \right) , \quad 
\varphi = \frac{r_- - r_+}{\ell} \left( \frac{t}{\ell} + \phi  \right)  \, ,
\end{eqnarray}
with $r_+$ and $r_-$ being the outer and inner horizons respectively and $\ell^2 = 8 \left(p^1\right)^2$ being the square of the radius of $AdS$, to rewrite the metric \eqref{metBTZuplift} as
\begin{eqnarray}\label{metBTZ}
 {\rm d}s^2_{(5)}   &=& 
 \left( - \frac{(r^2-r_+^2)(r^2-r_-^2)}{\ell^2 r^2 } {\rm d}t^2  + \frac{\ell^2 r^2 {\rm d}r^2}{(r^2-r_+^2)(r^2-r_-^2)}  + r^2 \left( {\rm d} \phi - \frac{r_+ r_-}{\ell r^2} {\rm d}t \right)^{2} \right) \nonumber \\
 &&\qquad  +2(p^1)^2 \, {\rm d}\Omega_{(2)}^2 \; .
\end{eqnarray}
The contribution in brackets can now be recognised as the standard non-extremal, rotating BTZ metric with radius fixed by $\ell$, mass and angular momentum given by
\begin{eqnarray}
M_{BTZ} = \frac{r_+^2 + r_-^2}{\ell^2} \; , \qquad 
J_{BTZ} = \frac{2 r_+ r_-}{\ell} .
\end{eqnarray}
Our results are consistent with the general black string solutions discussed in \cite{Compere:2010fm}. The total entropy of the uplifted, five-dimensional solution is now
\begin{align}
S ~ &= ~ \dfrac{1}{4G_5} ~ \mathrm{Area}(S^2) ~ 2 \pi r_+ \nonumber \\
&= ~ \dfrac{1}{4G_3} ~ 2 \pi r_+.
\end{align}
The metric \eqref{metBTZ} can be written entirely in terms of $M_{BTZ}$ and $J_{BTZ}$ as 
\begin{eqnarray}
 {\rm d}s^2_{(5)} ~ = ~ 
 \left( - f(r) ~ {\rm d}t^2  + f^{-1}(r) {\rm d}r^2  + r^2 \left( {\rm d} \phi + N_\phi {\rm d}t \right)^{2} \right) ~ + ~ 2(p^1)^2 \, {\rm d}\Omega_{(2)}^2 \; ,
\end{eqnarray}
where 
\begin{equation}
f(r) ~ = ~ - M_{BTZ} + \frac{r^2}{\ell^2} + \frac{J^2_{BTZ}}{4 r^2} \qquad N_\phi ~ = ~ - \frac{J_{BTZ}}{2 r^2}.
\end{equation}
This is written in conventions where the $AdS_3$ mass is $-1$, as opposed to $-\frac{1}{8G_3}$. One may restore the factors of $G_3$ by 
\begin{equation}
f(r) ~ = ~ - 8 G_3 M_{BTZ} + \frac{r^2}{\ell^2} + \frac{16 G^2_3 J^2_{BTZ}}{ r^2}, \qquad N_\phi ~ = ~ - \frac{4 G_3 J_{BTZ}}{r^2}\ ,
\end{equation}
and is now identical (up to a shift in the radial variable) to the metric written in  \cite{Strominger:1997eq}.\\

Given that the BTZ geometry arises in the bulk supergravity, following the results of \cite{Strominger:1997eq} and \cite{Brown:1986nw}, it is clear that the entropies of the macroscopic solution and the microscopic field theory match with each other. In fact, the central charges $c_L$ and $c_R$ of the CFT do not feel the boundary conditions, so they can be used again in the Cardy formula. However, the conventional argument\textemdash in \cite{Maldacena:1997de,Strominger:1996sh}, for instance\textemdash is that given a macroscopic black hole with certain (electric) charge, one may choose a conformal field theory with states carrying the same momentum that reproduces the macroscopic entropy. It is crucial, therefore, that the quantization conditions on the black hole charge and the field theory momentum are the same. In the case of supersymmetric black holes, both were integers and consequently consistent with each other. We saw in \eqref{eqn:quantizationinrho} that the momentum along the string is quantized; this becomes the four-dimensional electric charge, 
\begin{equation}
q_0=n_L-\left(N_R+\frac{c_R}{6}\left(\frac{\rho}{2}-\frac{\rho^2}{4}\right)\right)\ .
\end{equation}
To leading order in $g=\rho/2$, this is $a + b\, \frac{\rho}{2} = a + b\, g$, where $a$ and $b$ are integers\footnote{As shown in \cite{Sevrin:1988ew}, $\frac{c_R}{6}$ is an integer.}. Therefore, it is important that our macroscopic black hole satisfies this condition. We will now present a quick argument why the black hole \eqref{non-extr-r-coord} does satisfy this quantization condition. \\

For the black hole under consideration to be a physically reasonable one, it needs to have been formed by a collapse of particles within the theory. Elementary zero-mode particles in our theory have charges proportional to the gauge coupling constant\footnote{The factors of $R^3$ in the charges arise from issues of canonical normalization of the four-dimensional vectors. These are clearly not present in the five-dimensional `normalizations'.} $g$. The most general black hole in this theory could conceivably be formed by a collapse of Kaluza-Klein particles with integer charges and Scherk-Schwarz particles with charges proportional to $g$. Picking a black hole formed by $n_L - N_R$ Kaluza-Klein particles and $\frac{c_R}{6}$ Scherk-Schwarz particles, it has an electric charge that is exactly consistent with the quantization condition on the microscopic momentum \eqref{eqn:quantizationinrho}, to leading order in $\rho$. It would be interesting to understand the macroscopic origin of the term in \eqref{eqn:quantizationinrho} that is quadratic in $g$. For black hole temperatures larger than the supersymmetry breaking scale, this term is irrelevant. For lower temperatures, this correction can perhaps be understood after integrating out the hypermultiplets in a one-loop approximation. We leave this interesting point for future work. \\

\section{Extensions to supersymmetric vacua}\label{sec:possdir}

The construction we have presented so far needs attention to one further detail. We have considered a non-extremal black hole in a vacuum that spontaneously breaks supersymmetry. It is, therefore, important that the vacuum is at least sufficiently stable to allow for the formation of such a large black hole.To avoid possible problems with instabilities, we now present an alternative example in which supersymmetry is only partially broken in the vacuum. Since the discussion is very similar to the previous section, we will be rather brief and sketchy, only concentrating on the main steps. \\

Let us consider Type IIB Superstring theory on a $K3$ surface, preserving sixteen supercharges. This yields a six dimensional chiral $(0,4)$ supergravity theory supplemented with a moduli space, parametrized by the scalar fields,
\begin{equation}
\mathcal{M} ~ = ~ \dfrac{SO(5,21)}{SO(5)_R \times SO(21)}\ ,
\end{equation} 
where the $SO(5)_R\simeq USp(4)_R$ is the R-symmetry. This R-symmetry group contains two compact $U(1)$ subgroups, labelled by say, $U(1)_{\rho_L}$ and $U(1)_{\rho_R}$, 
\begin{equation}
U(1)_{\rho_L}\times U(1)_{\rho_R} \subset SO(5)_R\ .
\end{equation}

One may now repeat the construction we have presented in this article, and compactify further on a circle with a Scherk-Schwarz twist, this time down to five dimensions. This procedure leads to a Scherk Schwarz reduced gauged $\mathcal{N} = 4$ supergravity in five dimensions. For toroidal compactifications that result in maximal supersymmetry in six dimensions, such partial supersymmetry breaking flat vacua have been shown to exist\cite{Andrianopoli:2004xu}. For theories arising from K3 compactifications, a similar feature has been shown in \cite{Villadoro:2004ci}. Applied to the case at hand, one can twist the six-dimensional supercharges with respect to $U(1)_{\rho_L}\times U(1)_{\rho_R} \subset SO(5)_R$, with twist parameters $\rho_L$ and $\rho_R$. If both parameters are switched on, supersymmetry is completely broken in the vacuum. However, if we set, say $\rho_L=0$, supersymmetry is only partly broken\textemdash and two of four gravitinos remain massless:
\begin{equation}
SO(5)_R \simeq USp(4)_R \longrightarrow USp(2)_R\simeq SU(2)_R\ .
\end{equation}
Further details on the spectrum can be found in \cite{Villadoro:2004ci}. \\

Therefore, setting $\rho_L = 0$ leaves us with an $\mathcal{N} = 2$ preserving Minkowski vacuum in five dimensions. Given that a stable vacuum is now guaranteed, it is no longer problematic to consider a non-extremal black hole excitation above this vacuum. In fact, one may even stick to the extremal case. Following up on the spectrum computed in \cite{Villadoro:2004ci}, for example, it is straightforward to check that the appropriate quantization condition on the electric charge of these black holes is consistent with the expectation from the $\rho$-algebras.\\

In such a set up, an extension of the Rholographic picture is simple too. A black string solution of the six dimensional supergravity theory has an $AdS_3 \times S^3$ horizon. In fact, this was the set up considered in the classic example of \cite{Strominger:1996sh}. Its Rholographic counterpart would be the $\rho_{L/R}$-twisted non-extremal excitation on the $AdS^{\rho_{L/R}}_3$ vacuum. The field theory living on its boundary is a $(4,4)$ D1-D5 CFT. It contains two chiral $\mathcal{N}=4$ superconformal algebras in two dimensions. For the Rholographic extension of which, as discussed at the end of Section \ref{sec:rhoalgebras}, one may consider a $\rho_{L/R}$-algebra extension on either of the chiral components of this CFT. Therefore, a $\rho_{L/R}$-twisted D1-D5 CFT is conjecturally dual to Type IIB Superstring theory on an $AdS^{\rho_{L/R}}_3 \times S^3 \times K3$. \\

It is worth noting that the D1-D5 CFT has local gauge symmetry that leads to spectral-flow, much like in the case of the MSW CFT. While there was one set of Kac-Moody currents corresponding to the $SU(2)$ gauge symmetry in the MSW CFT (corresponding to rotational symmetry on the $S^2$ of the $AdS_3$ horizon), the D1-D5 CFT has two such current algebras corresponding to rotational symmetry on the $S^3$, with an isometry group $SO(4) \simeq SU(2) \times SU(2)$. It must be stressed that the Scherk-Schwarz twist on the worldsheet does not involve the current algebras. Rather, it uses the outer automorphism groups of the left and right moving sectors, which we call $SU(2)_{\rho_L}\times SU(2)_{\rho_R}$. It is clear then that the twists on the worldsheet supercharges is with respect to the subgroups
\begin{equation}
U(1)_{\rho_L}\times U(1)_{\rho_R}\subset SU(2)_{\rho_L}\times SU(2)_{\rho_R}\ ,
\end{equation}
and if we want to preserve some supersymmetry in the bulk, we set one of the twist parameters to zero, e.g. $\rho_L=0$. The concerned reader may consider this example to be on more firm ground, as far as stability of the vacuum is concerned. In fact, it would be interesting to compute black hole discharge rates and R-charged particle scattering processes using conformal field theory techniques for the $\rho$-algebras. It would also be interesting to explore the consequences of, and find more evidence for, the Rholographic picture. We leave these interesting questions for future research.

\section*{Acknowledgements}

It is a pleasure to thank Jan de Boer, Clay Cordova, Chris Hull, Elias Kiritsis, Finn Larsen and Cumrun Vafa for interesting comments and/or stimulating discussions. 
This work was supported by the Netherlands Organisation for Scientific Research (NWO) under the VICI grant 680-47-603, and the Delta-Institute for Theoretical Physics (D-ITP) that is funded by the Dutch Ministry of Education, Culture and Science (OCW). This work was also supported by the Interuniversity Attraction Poles Programme initiated by the Belgian Science Policy (P7/37), and by COST Action MP1210 The String Theory Universe. O.V. is supported by a Marie Curie fellowship and is grateful to the CPHT of  Ecole Polytechnique for managing the administration of this grant. O.V. is also supported in part by DOE grant de-sc0007870.

\appendix

\section{Compact gauging of G$_{2(2)}$/SO(4)} \label{app:KVs}

Here we identify the relevant $U(1)_R$ of the model discussed in the main text and then compute its associated Killing vector and moment map. Let $H_1$, $H_2$ be the two Cartans and $E_i$,  $F_i$, $i=1, \ldots, 6$, the positive and negative root generators of the split real form G$_{2(2)}$. The maximally compact subgroup SO(4) is generated by
\begin{eqnarray}\label{eqn:G2killings1}
&& K_1 = E_1 - \delta^2 \gamma^{-2} \, F_1 \; , \qquad 
K_2 = E_2 - \gamma^{2} \, F_2 \; , \qquad 
K_3 = E_3 - \delta^2  \, F_3 \; , \nonumber \\
&& K_4 = E_4 - \gamma^2 \delta^2 \, F_4 \; , \qquad 
K_5 = E_5 - \gamma^{4} \delta^2 \, F_5 \; , \qquad 
K_6 = E_6 - \gamma^2 \delta^4  \, F_6 \; , \nonumber \\
\end{eqnarray}
for any non-zero real constants $\gamma$ and $\delta$. Indeed, the further combinations
\begin{eqnarray}\label{eqn:G2killings2}
J_1 = \tfrac12 \delta^{-1} \big( \gamma^{-2}  K_5 - \sqrt{3} K_3 \big) \; , \quad 
J_2 = \tfrac12 \gamma^{-1}  \big( \delta^{-2}  K_6 + \sqrt{3} K_2 \big) \; , \quad 
J_3 = \tfrac12 \gamma  \delta^{-1}  \big( K_1 - \gamma^{-2}  \sqrt{3} K_4 \big) \; , \nonumber \\
\end{eqnarray}
and
\begin{eqnarray}
L_1 = \tfrac12  \delta^{-1}  \big( 3  \gamma^{-2}  K_5 + \sqrt{3} K_3 \big) \; , \;
L_2 = \tfrac12 \gamma^{-1}  \big( 3 \delta^{-2} K_6 - \sqrt{3} K_2 \big) \; , \;
L_3 = \tfrac12  \gamma  \delta^{-1}   \big( 3K_1 + \gamma^{-2}  \sqrt{3} K_4 \big) \; ,  \nonumber \\
\end{eqnarray}
can be checked to generate two copies of SU(2), for any $\gamma$ and $\delta$. This is most straightforwardly seen using an explicit matrix realisation of the G$_{2(2)}$ generators, like {\it e.g.} the one given in appendix C of \cite{Donos:2010ax}.  A calculation similar to that of that appendix allows us to establish that the SU(2)$_R \approx$Sp(1) corresponding to the R-symmetry is generated by  $J_x$, $x=1,2,3$. Any of the $J_x$ can thus be picked up as the relevant U$(1)_R$ to gauge our model with.  For definiteness, we choose\footnote{We have explicitly verified that graviphoton gaugings along $J_1$ only, along $J_2$ only or along $J_3$ only are physically indistinguishable, as they should.}  $J_3$. \\

We now turn to the calculation of the Killing vector associated to $J_3$. The Killing vectors of hypermultiplet spaces in the image of the c-map have been given in terms of special geometry data in \cite{deWit:1992wf} (see \cite{Erbin:2014hsa} for a recent update). Here, rather than using those general formulae, we play the following trick, based on the homogeneity of G$_{2(2)}$/SO(4), to read off the Killing vector associated to a specific generator. If ${\cal V} (q^u)$ is the right, say, coset representative and $\sharp$ denotes the G$_{2(2)}$-generalised transpose (see e.g. \cite{Donos:2010ax} for the details), then
 $P= \frac12 \left( d{\cal V} \, {\cal V}^{-1} + \left( d{\cal V} \, {\cal V}^{-1}\right)^\sharp \right)$ 
 is a one-form valued on the Lie algebra   $g_{2(2)}$ of  $G_{2(2)}$. For any real one-form $A$, the one-form
  $\hat P= \frac12 \left( D{\cal V} \, {\cal V}^{-1} + \left( D{\cal V} \, {\cal V}^{-1}\right)^\sharp \right)$, 
 with
  $D{\cal V} \, {\cal V}^{-1} \equiv (d{\cal V} + g A \, {\cal V} \, J_3 ) {\cal V}^{-1} $, is also $g_{2(2)}$-valued. Here we have  found it useful to stick in a (coupling) constant $g$. We can then expand   $\hat P$ in the basis $H_1$, $H_2$, $E_i$,  $F_i$ of $g_{2(2)}$ to read off the covariant derivative $Dq^u = d q^u + g A \, k^u$ and the components of the $k^u$ of the Killing vector associated to the generator $J_3$. In fact, we have repeated this exercise for all 14 generators of G$_{2(2)}$ to compute all Killing vectors of G$_{2(2)}$/SO(4), and have explicitly verified that these vectors do indeed leave the metric $h_{uv} dq^u dq^v = \frac14 \mathrm{Tr} (PP)$ invariant. Obviously,  the same process can be followed to compute the Killing vectors of any (non-compact) homogeneous space. \\

Performing the suitable coordinate transformation that brings the metric   $h_{uv} dq^u dq^v = \frac14 \mathrm{Tr} (PP)$  obtained from the coset approach into the c-map form \eqref{cmapmetricG2SO4}, we thus find that the Killing vector $k_0 = k_0^u \, \partial_u$ associated to the $U(1)_R$ generator $J_3$  has the following components $k_0^u$ along the coordinates \eqref{eq:G2SO4Coords}:
\begin{eqnarray} \label{KillR}
k_0^\phi & = & -2^{-\frac32} \, 3^{-\frac34}\, \gamma^{-1} \delta \, \big( \xi^0 + 3\sqrt{3} \, \delta^2\,  \tilde\xi_1 \big)  \; , \nonumber \\[6pt]
k_0^\varphi & = & 2^{-\frac32} \, 3^{-\frac34} \, \gamma^{-1} \delta \, \big( \xi^0 -4 \sqrt{3}  \, \gamma^2 \,  \chi \,  \xi^1  -\sqrt{3} \,  \gamma^2\, \tilde\xi_1 \big)   \; , \nonumber \\[6pt]
k_0^\chi & = & -2^{-\frac12} \, 3^{-\frac54} \, \gamma^{-1} \delta \,  \Big(  \sqrt{3} \, \chi \, \xi^0 - \big(1-6 \gamma^2 e^{-4\varphi} + 6 \gamma^2 \chi^2 \big) \, \xi^1 +3\sqrt{3} \, \gamma^2\, \tilde\xi_0 - 3 \chi \, \gamma^2 \, \tilde \xi_1 \Big)   \; , \nonumber \\[6pt]
k_0^a & = & 2^{-\frac32} \, 3^{-\frac{11}{4}}  \, \gamma^{-1} \delta^{-1} \,  \Big( 9 \delta^2 a \big( \xi^0 +3\sqrt{3} \,  \gamma^2 \, \tilde\xi_1 \big) 
-9 \delta^2 e^{-2\phi-2\varphi}  \chi \,  \xi^0 \big[ 9 \gamma^2 + 18 \gamma^2 e^{4\varphi} \chi^2 -e^{8\varphi} \chi^2 (1-9\gamma^2 \chi^2) \big]  \nonumber \\
&& \qquad \qquad \quad +9\sqrt{3}  e^{-2\phi-2\varphi} \, \xi^1 \big[ 3 \gamma^2 \delta^2 + \sqrt{3} e^{2\phi +2\varphi} +12 \gamma^2 \delta^2 e^{4\varphi} \chi^2 -\delta^2 e^{8\varphi} \chi^2 (1-9\chi^2) \big]  \nonumber \\
&& \qquad \qquad \quad -9\, \tilde \xi_0 \big[ 3\sqrt{3} \gamma^2 + \delta^2 e^{-2\phi +6\varphi}  (1-9\chi^2) \big]  
+ 9\sqrt{3} \, \delta^2 e^{-2\phi+2\varphi}  \chi \,  \tilde \xi_1  \big[ 6\gamma^2 - e^{4\varphi}  (1-9\gamma^2\chi^2) \big]  \nonumber \\
&& \qquad \qquad \quad +9 \delta^2 \xi^0 \big[ \xi^0 \tilde\xi_0 + \xi^1 \tilde\xi_1 -3\sqrt{3} \, \gamma^2\, \tilde\xi_0 \,\tilde\xi_1  \big]  -9\delta^2 \xi^1 \big[ 54 \gamma^2 \xi^0 \tilde\xi_0 -2\sqrt{3} (\xi^1)^2 -9\sqrt{3} \, \gamma^2(\tilde{\xi}_1)^2  \big]  \Big)   \; , \nonumber \\[6pt]
k_0^{\xi^0} & = & 2^{-\frac32} \, 3^{-\frac34} \, \gamma^{-1} \delta^{-1} \,  \Big( 2 \delta^2 (\xi^0)^2  -6\gamma^2 \delta^2 (\xi^1)^2 +3 \sqrt{3} \, \gamma^2 - \delta^2 e^{-2\phi +6\varphi}  (1-9\gamma^2 \chi^2)  \Big)   \; , \nonumber \\[6pt]
k_0^{\xi^1}    & = & -2^{-\frac32} \, 3^{-\frac34} \, \gamma^{-1} \delta \,  \Big( 3\sqrt{3} \, \gamma^2 \, a -2 \xi^0 \xi^1 + 3\sqrt{3}  \, \gamma^2 \,   \xi^0  \tilde \xi_0  -5\sqrt{3} \, \gamma^2 \, \xi^1 \tilde\xi_1  -6\sqrt{3} \,  \gamma^2 \, \chi e^{-2\phi +2\varphi}  \nonumber \\
&& \qquad \qquad\qquad \qquad +\sqrt{3} e^{-2\phi +6\varphi} \chi (1-9 \gamma^2 \chi^2)  \Big)   \; , \nonumber \\[6pt]
k_0^{\tilde{\xi_0}} & = & 2^{-\frac32} \, 3^{-\frac34}  \, \gamma^{-1} \delta \,  \Big( a - \xi^0 \tilde\xi_0 - \xi^1  \tilde \xi_1  +6\sqrt{3} \, \gamma^2 \, \tilde\xi_0 \tilde\xi_1 +9\gamma^2 \chi e^{-2\phi -2\varphi}  +18 \gamma^2 e^{-2\phi +2\varphi} \chi^3 \nonumber \\
&& \qquad \qquad\qquad \qquad - e^{-2\phi +6\varphi} \chi^3  (1-9 \gamma^2 \chi^2)  \Big)   \; , \nonumber \\[6pt]
k_0^{\tilde{\xi_1}} & = & -2^{-\frac32} \, 3^{-\frac54}  \, \gamma^{-1} \delta^{-1} \, \Big( 2 \delta^2 (\xi^1)^2 -12\sqrt{3}\, \gamma^2 \delta^2 \,  \xi^1 \tilde\xi_0 - 6 \, \gamma^2 \delta^2 \, (\tilde \xi_1)^2 -3\sqrt{3} + 9 \, \gamma^2 \delta^2 \, e^{-2\phi -2\varphi}  \nonumber \\
&& \qquad \qquad\qquad \qquad \quad  +36 \, \gamma^2 \delta^2 \, e^{-2\phi +2\varphi} \chi^2 - 3 \, \delta^2 \, e^{-2\phi +6\varphi} \chi^2  (1-9\gamma^2 \chi^2)  \Big)   \; .  
\end{eqnarray}
It is now straightforward to doublecheck by standard methods that  this vector leaves the metric (\ref{cmapmetricG2SO4}) invariant, and thus is indeed Killing, and that it vanishes at \eqref{VanishingPoint}. \\

We have also computed the moment map $P^x_0$, $x=1,2,3$, corresponding to this isometry. Since the full expression is not very illuminating, we only give its value at the vacuum  \eqref{VanishingPoint}, which is the only quantity needed for all our analyses. With the normalisation of \cite{Andrianopoli:1996cm}, we obtain
\begin{eqnarray}
P_0^x=(0,0,2) \ ,
\end{eqnarray}
independent of $\gamma$ and $\delta$. Since the moment map is independent of these factors, so is the mass (charge) spectrum.

\bibliographystyle{IEEEtran}
\bibliography{biblio-rholography}

\begin{thebibliography}{10}
\providecommand{\url}[1]{#1}
\csname url@samestyle\endcsname
\providecommand{\newblock}{\relax}
\providecommand{\bibinfo}[2]{#2}
\providecommand{\BIBentrySTDinterwordspacing}{\spaceskip=0pt\relax}
\providecommand{\BIBentryALTinterwordstretchfactor}{4}
\providecommand{\BIBentryALTinterwordspacing}{\spaceskip=\fontdimen2\font plus
\BIBentryALTinterwordstretchfactor\fontdimen3\font minus
  \fontdimen4\font\relax}
\providecommand{\BIBforeignlanguage}[2]{{%
\expandafter\ifx\csname l@#1\endcsname\relax
\typeout{** WARNING: IEEEtran.bst: No hyphenation pattern has been}%
\typeout{** loaded for the language `#1'. Using the pattern for}%
\typeout{** the default language instead.}%
\else
\language=\csname l@#1\endcsname
\fi
#2}}
\providecommand{\BIBdecl}{\relax}
\BIBdecl

\bibitem{Strominger:1996sh}
A.~Strominger and C.~Vafa, ``{Microscopic origin of the Bekenstein-Hawking
  entropy},'' \emph{Phys.Lett.}, vol. B379, pp. 99--104, 1996.

\bibitem{Maldacena:1997de}
J.~M. Maldacena, A.~Strominger, and E.~Witten, ``{Black hole entropy in M
  theory},'' \emph{JHEP}, vol. 9712, p. 002, 1997.

\bibitem{Callan:1996dv}
C.~G. Callan and J.~M. Maldacena, ``{D-brane approach to black hole quantum
  mechanics},'' \emph{Nucl.Phys.}, vol. B472, pp. 591--610, 1996.

\bibitem{Horowitz:1996fn}
G.~T. Horowitz and A.~Strominger, ``{Counting states of near extremal black
  holes},'' \emph{Phys.Rev.Lett.}, vol.~77, pp. 2368--2371, 1996.

\bibitem{Breckenridge:1996sn}
J.~Breckenridge, D.~Lowe, R.~C. Myers, A.~Peet, A.~Strominger \emph{et~al.},
  ``{Macroscopic and microscopic entropy of near extremal spinning black
  holes},'' \emph{Phys.Lett.}, vol. B381, pp. 423--426, 1996.

\bibitem{Dabholkar:1997rk}
A.~Dabholkar, ``{Microstates of nonsupersymmetric black holes},''
  \emph{Phys.Lett.}, vol. B402, pp. 53--58, 1997.

\bibitem{Sen:2007qy}
A.~Sen, ``{Black Hole Entropy Function, Attractors and Precision Counting of
  Microstates},'' \emph{Gen.Rel.Grav.}, vol.~40, pp. 2249--2431, 2008.

\bibitem{Strominger:1997eq}
A.~Strominger, ``{Black hole entropy from near horizon microstates},''
  \emph{JHEP}, vol. 9802, p. 009, 1998.

\bibitem{Schwimmer:1986mf}
A.~Schwimmer and N.~Seiberg, ``{Comments on the N=2, N=3, N=4 Superconformal
  Algebras in Two-Dimensions},'' \emph{Phys.Lett.}, vol. B184, p. 191, 1987.

\bibitem{Minasian:1999qn}
R.~Minasian, G.~W. Moore, and D.~Tsimpis, ``{Calabi-Yau black holes and (0,4)
  sigma models},'' \emph{Commun.Math.Phys.}, vol. 209, pp. 325--352, 2000.

\bibitem{Andrianopoli:2004im}
L.~Andrianopoli, S.~Ferrara, and M.~Lledo, ``{Scherk-Schwarz reduction of D = 5
  special and quaternionic geometry},'' \emph{Class.Quant.Grav.}, vol.~21, pp.
  4677--4696, 2004.

\bibitem{Looyestijn:2010pb}
H.~Looyestijn, E.~Plauschinn, and S.~Vandoren, ``{New potentials from
  Scherk-Schwarz reductions},'' \emph{JHEP}, vol. 1012, p. 016, 2010.

\bibitem{Scherk:1978ta}
J.~Scherk and J.~H. Schwarz, ``{Spontaneous Breaking of Supersymmetry Through
  Dimensional Reduction},'' \emph{Phys.Lett.}, vol. B82, p.~60, 1979.

\bibitem{Gaiotto:2006wm}
D.~Gaiotto, A.~Strominger, and X.~Yin, ``{The M5-Brane Elliptic Genus:
  Modularity and BPS States},'' \emph{JHEP}, vol. 0708, p. 070, 2007.

\bibitem{Intriligator:1998ig}
K.~A. Intriligator, ``{Bonus symmetries of N=4 superYang-Mills correlation
  functions via AdS duality},'' \emph{Nucl.Phys.}, vol. B551, pp. 575--600,
  1999.

\bibitem{Yu:1987dh}
M.~Yu, ``{The Unitary Representations of the $N=4$ SU(2) Extended
  Superconformal Algebras},'' \emph{Nucl.Phys.}, vol. B294, p. 890, 1987.

\bibitem{Defever:1988um}
F.~Defever, S.~Schrans, and K.~Thielmans, ``{Moding of Superconformal
  Algebras},'' \emph{Phys.Lett.}, vol. B212, p. 467, 1988.

\bibitem{Henneaux:1999ib}
M.~Henneaux, L.~Maoz, and A.~Schwimmer, ``{Asymptotic dynamics and asymptotic
  symmetries of three-dimensional extended AdS supergravity},'' \emph{Annals
  Phys.}, vol. 282, pp. 31--66, 2000.

\bibitem{Braun:2011hd}
V.~Braun, ``{The 24-Cell and Calabi-Yau Threefolds with Hodge Numbers (1,1)},''
  \emph{JHEP}, vol. 1205, p. 101, 2012.

\bibitem{Cadavid:1995bk}
A.~Cadavid, A.~Ceresole, R.~D'Auria, and S.~Ferrara, ``{Eleven-dimensional
  supergravity compactified on Calabi-Yau threefolds},'' \emph{Phys.Lett.},
  vol. B357, pp. 76--80, 1995.

\bibitem{deWit:1998zg}
B.~de~Wit, B.~Kleijn, and S.~Vandoren, ``{Rigid N=2 superconformal
  hypermultiplets},'' \emph{Lect.Notes Phys.}, vol. 524, p.~37, 1999.

\bibitem{deWit:1999fp}
------, ``{Superconformal hypermultiplets},'' \emph{Nucl.Phys.}, vol. B568, pp.
  475--502, 2000.

\bibitem{Scherk:1979zr}
J.~Scherk and J.~H. Schwarz, ``{How to Get Masses from Extra Dimensions},''
  \emph{Nucl.Phys.}, vol. B153, pp. 61--88, 1979.

\bibitem{Bergshoeff:1997mg}
E.~Bergshoeff, M.~de~Roo, and E.~Eyras, ``{Gauged supergravity from dimensional
  reduction},'' \emph{Phys.Lett.}, vol. B413, pp. 70--78, 1997.

\bibitem{Dabholkar:2002sy}
A.~Dabholkar and C.~Hull, ``{Duality twists, orbifolds, and fluxes},''
  \emph{JHEP}, vol. 0309, p. 054, 2003.

\bibitem{Andrianopoli:1996cm}
L.~Andrianopoli, M.~Bertolini, A.~Ceresole, R.~D'Auria, S.~Ferrara
  \emph{et~al.}, ``{N=2 supergravity and N=2 superYang-Mills theory on general
  scalar manifolds: Symplectic covariance, gaugings and the momentum map},''
  \emph{J.Geom.Phys.}, vol.~23, pp. 111--189, 1997.

\bibitem{Bodner:1989cg}
M.~Bodner and A.~Cadavid, ``{Dimensional Reduction of Type IIb Supergravity and
  Exceptional Quaternionic Manifolds},'' \emph{Class.Quant.Grav.}, vol.~7, p.
  829, 1990.

\bibitem{Ferrara:1989ik}
S.~Ferrara and S.~Sabharwal, ``{Quaternionic Manifolds for Type II Superstring
  Vacua of Calabi-Yau Spaces},'' \emph{Nucl.Phys.}, vol. B332, p. 317, 1990.

\bibitem{Cassani:2012pj}
D.~Cassani, P.~Koerber, and O.~Varela, ``{All homogeneous N=2 M-theory
  truncations with supersymmetric AdS4 vacua},'' \emph{JHEP}, vol. 1211, p.
  173, 2012.

\bibitem{Hristov:2012nu}
K.~Hristov, S.~Katmadas, and V.~Pozzoli, ``{Ungauging black holes and hidden
  supercharges},'' \emph{JHEP}, vol. 1301, p. 110, 2013.

\bibitem{Hristov:2014eba}
K.~Hristov and A.~Rota, ``{6d-5d-4d reduction of BPS attractors and black
  objects in flat gauged supergravities},'' 2014.

\bibitem{Hristov:2010eu}
K.~Hristov, H.~Looyestijn, and S.~Vandoren, ``{BPS black holes in N=2 D=4
  gauged supergravities},'' \emph{JHEP}, vol. 1008, p. 103, 2010.

\bibitem{Galli:2011fq}
P.~Galli, T.~Ortin, J.~Perz, and C.~S. Shahbazi, ``{Non-extremal black holes of
  N=2, d=4 supergravity},'' \emph{JHEP}, vol. 1107, p. 041, 2011.

\bibitem{Chen:2010bsa}
C.-M. Chen, Y.-M. Huang, and S.-J. Zou, ``{Holographic Duals of Near-extremal
  Reissner-Nordstrom Black Holes},'' \emph{JHEP}, vol. 1003, p. 123, 2010.

\bibitem{Compere:2010fm}
G.~Compere, S.~de~Buyl, S.~Stotyn, and A.~Virmani, ``{A General Black String
  and its Microscopics},'' \emph{JHEP}, vol. 1011, p. 133, 2010.

\bibitem{Brown:1986nw}
J.~D. Brown and M.~Henneaux, ``{Central Charges in the Canonical Realization of
  Asymptotic Symmetries: An Example from Three-Dimensional Gravity},''
  \emph{Commun.Math.Phys.}, vol. 104, pp. 207--226, 1986.

\bibitem{Sevrin:1988ew}
A.~Sevrin, W.~Troost, and A.~Van~Proeyen, ``{Superconformal Algebras in
  Two-Dimensions with N=4},'' \emph{Phys.Lett.}, vol. B208, p. 447, 1988.

\bibitem{Andrianopoli:2004xu}
L.~Andrianopoli, S.~Ferrara, and M.~Lledo, ``{No-scale D=5 supergravity from
  Scherk-Schwarz reduction of D=6 theories},'' \emph{JHEP}, vol. 0406, p. 018,
  2004.

\bibitem{Villadoro:2004ci}
G.~Villadoro and F.~Zwirner, ``{The Minimal N=4 no-scale model from generalized
  dimensional reduction},'' \emph{JHEP}, vol. 0407, p. 055, 2004.

\bibitem{Donos:2010ax}
A.~Donos, J.~P. Gauntlett, N.~Kim, and O.~Varela, ``{Wrapped M5-branes,
  consistent truncations and AdS/CMT},'' \emph{JHEP}, vol. 1012, p. 003, 2010.

\bibitem{deWit:1992wf}
B.~de~Wit, F.~Vanderseypen, and A.~Van~Proeyen, ``{Symmetry structure of
  special geometries},'' \emph{Nucl.Phys.}, vol. B400, pp. 463--524, 1993.

\bibitem{Erbin:2014hsa}
H.~Erbin and N.~Halmagyi, ``{Abelian Hypermultiplet Gaugings and BPS Vacua in N
  = 2 Supergravity},'' 2014.

\end{thebibliography}

\end{document}